\documentclass[aps,prd,twocolumn,article]{revtex4}
\usepackage{ulem}
\usepackage{color}
\usepackage{graphicx}
\usepackage{epsfig}
\usepackage{amsmath} 
\usepackage{tikz} 
\usepackage{comment}
\usepackage{natbib}

\usepackage{bm}
\newcommand{\be}{\begin{equation}}
\newcommand{\ee}{\end{equation}}
\newcommand{\ba}{\begin{eqnarray}}
\newcommand{\ea}{\end{eqnarray}}

\begin{document}
\title{Impact of off-shell dynamics on the transport properties and the dynamical evolution of Charm Quarks at RHIC and LHC temperatures } 
\author{Maria Lucia Sambataro$^{a,b}$, Salvatore Plumari$^{a,b}$, Vincenzo Greco $^{a,b}$}

\affiliation{$^a$ Department of Physics and Astronomy 'E. Majorana', University of Catania, 
Via S. Sofia 64, 1-95125 Catania, Italy}
\affiliation{$^b$ Laboratori Nazionali del Sud, INFN-LNS, Via S. Sofia 62, I-95123 Catania, Italy}
\date{\today}

\begin{abstract}
We evaluate drag and diffusion transport coefficients comparing a quasi-particle approximation
with on-shell constituents of the QGP medium and a dynamical quasi-particles model with off-shell
bulk medium at finite temperature T. We study the effects of the width $\gamma$ of the particles of
the bulk medium on the charm quark transport properties exploring the range where $\gamma < M_{q,g}$. 
We find that off-shell effects are in general quite moderate and can induce a reduction of the drag
coefficient at low momenta that disappear already at moderate  momenta, $p \gtrsim 2-3\, \rm GeV$. 
We also observe a moderate reduction of the breaking of the Fluctuation-Dissipation theorem (FDT)
at finite momenta.  

Moreover, we have performed a first study of the dynamical evolution of HQ elastic energy loss  
in a bulk medium at fixed temperature extending the Boltzmann (BM) collision integral to include
off-shell dynamics.
A comparison among the Langevin dynamics, the BM collisional integral with on-shell and
the BM extension to off-shell dynamics shows that the evolution of charm energy when off-shell effects are included
remain quite similar to the case of the on-shell BM collision integral.

\vspace{2mm}

\end{abstract}

\maketitle

\section{Introduction}

Heavy quarks (HQs), namely charm and bottom, are considered as a solid probe
to characterize the matter created in the QGP phase 
~\cite{Svetitsky:1987gq,Moore:2004tg,Dong:2019unq,Prino:2016cni,Andronic:2015wma,Aarts:2016hap,Cao:2018ews,Rapp:2018qla}.
The large mass of heavy quarks has several implications in this context.
They are produced in the early stage of the collisions by pQCD process and being
$M_{HQ}>> T$ also the thermal pair production and annihilation processes are negligible.
For a perturbative interaction due also to the large mass leading to collisions
with small momentum transfer, the
HQ propagation through the QGP medium can be described as a diffusion
process assimilated to a Brownian motion \cite{Moore:2004tg,Das:2013kea,Dong:2019unq}.
Furthermore, the large mass has the effect to reduce the equilibration rate of heavy quarks in the medium relative 
to their light counterparts leading to a thermalization time comparable to the one of the life time of the fireball
\cite{Greco:2017rro,Dong:2019unq}.
Therefore, the standard approach to describe the propagation of HQ in QGP has
been quite often treated within the framework of the Fokker-Planck equation 
\cite{Moore:2004tg,vanHees:2004gq,vanHees:2005wb,vanHees:2007me,LV1,Qin:2010pf,Das:2012ck,
Das:2010tj,Alberico:2013bza,He:2011qa,He:2014cla,Xu:2017obm}. However,
the evidence of non-perturbative interaction and the large initial temperatures
at LHC, $M_c \simeq 3 T \simeq <p_{bulk}>$, hint at a scattering dynamics more appropriately described by a Boltzmann
collision integral that implies significant deviations from a gaussian fluctuation around the
average momentum of the charm quark \cite{Das:2013kea,Gossiaux:2008jv,Gossiaux:2009mk,Ghosh:2011bw,Uphoff:2012gb,Uphoff:2011ad,Song:2015sfa,Cao:2016gvr,Cao:2017hhk}.

One of the main observable for HQs that it has been also extensively used
as a probe of QGP, is the nuclear suppression factor, $R_{AA}(p_T)$,
~\cite{Abelev:2006db,Adam:2015sza,Adler:2005xv}. It is defined as
the ratio  between the heavy flavor hadrons produced in nucleus-nucleus collisions with
respect to those produced in proton-proton  collisions.
Another observable extensively studied is the elliptic flow~\cite{Adare:2006nq,Abelev:2014ipa}, $v_2(p_T)=<cos(2\phi)>$, a measure of the anisotropy in the  angular distribution of heavy mesons in momentum space, as a response to the initial anisotropy 
in coordinate space in non-central collisions.
In literature several studies have been performed in these years, to theoretically study both these observables with the aim to understand heavy quark dynamics in QGP employing the Langevin or the on-shell Boltzmann transport equation
~\cite{vanHees:2005wb,vanHees:2007me,Gossiaux:2008jv,Das:2009vy,Alberico:2011zy,Uphoff:2012gb,Lang:2012cx,Song:2015ykw,Das:2013kea,Cao:2015hia,Das:2015ana,Cao:2017hhk,Das:2017dsh,Sun:2019fud,rol2019size,Plumari:2019hzp}.
However, being the QGP strongly interacting, a full quantum description of the charm quark interaction
should include in principle also the off-shell dynamics, an approach that has been developed only in
\cite{Berrehrah:2013mu} for the study of the transport coefficients and it is included in the PHSD
approach to heavy-ion collisions \cite{Song:2015ykw,Song:2015sfa,Song:2016rzw}.
In this paper we extended this study exploring also the effects of larger widths and in particular
discussing the effect also in terms of the Fluctuations Dissipation Theorem (FDT).
Moreover, we present a first study of the time evolution
of the charm momentum in a bulk medium at fixed temperature T comparing directly
the Langevin evolution, the Boltzmann on-shell evolution and an extension of the Boltzmann
collision integral to include off-shell dynamics. We also discuss the impact that off-shell dynamics can
have on the $R_{AA}(p_T)$.

The paper is organized as follows. In the next sections we will briefly present the on-shell
Boltzmann transport equation, the Fokker-Planck (Langevin) one and the definition of drag and diffusion coefficients
in both on-shell and off-shell approaches.
In section \ref{Section:3}, we discuss the results obtained for the transport coefficients
 in both on-shell and off-shell models.
Section \ref{Section:4} is devoted to discuss the dynamical evolution of charm quarks in a bulk medium at finite T
by comparing  the results obtained in Langevin, on-shell and off-shell Boltzmann approaches.
Section \ref{Section:5} contains the summary and conclusions.

\section{Boltzmann transport equation and transport coefficients}
In this section we are interested to study both the transport coefficients and the time evolution of
the phase-space distribution function of heavy quarks (HQs). The starting point in the study of propagation 
of heavy quark is the relativistic transport equation for HQs scattering in a bulk medium of quarks and gluons.
We therefore briefly describe the relativistic Boltzmann-Vlasov equation from which we will deduce the
transport coefficients for on-shell dynamics and the Fokker-Planck equation.
The on-shell transport equation can be expressed by the Boltzmann-Vlasov equation
given by the following integro-differential equation: 
\begin{eqnarray}
\{p^\mu\partial_\mu+m^*(x)\partial_\mu m^*(x)\partial^{\mu}_p\}f_Q(x,p)=C[f_q,f_g,f_{Q}] \nonumber \\
\end{eqnarray}
where $f_Q(x,p)$ and $f_{q,g}(x,p)$ are the phase-space distribution functions for the heavy quark
and light quarks and gluons respectively, while $C[f_q, f_g, f_Q](x,p)$ is the relativistic
Boltzmann-like collision integral allowing to describe the short range interaction between
heavy quark and particles of plasma. The distribution function of the bulk medium of
quarks and gluons has in general to be determined by another set of equations
that could be the Boltzmann-Vlasov equations for quark and gluons
\cite{Ruggieri:2013ova,Plumari:2015cfa,Plumari:2019gwq}.
In the present study, we want to address a direct comparison between two different dynamics: the relativistic
Langevin dynamics and the relativistic Boltzmann transport theory. 
In the second approach, we will discuss the role of on-shell and off-shell effects on the HQ dynamics.
In order to have a better focusing and testing the dynamics between
these different approaches the bulk medium will be considered as a
thermal bath at equilibrium at some temperature $T$.
Moreover, we will calculate the different transport coefficients
of HQs in a static medium at finite temperature. This will give
the response of the medium to the propagation of HQs under
fixed thermodynamical conditions.
This is a first step before studying the more complex case of the expanding medium
in realistic uRHIC where gradients of density and temperature are involved.
Therefore in our calculations we neglect effects caused by 
space-time variation of the scalar mean fields, $\partial_\mu m^*(x)\approx 0$. 
Assuming that the distribution function is $x$ independent, i.e. the plasma
is uniform, each variation of the distribution function is due to collisions
and the Boltzmann equation is simplified to a integro-differential equation
only respect to time: 
\begin{equation}
\label{Boltz_simply}
p^0\partial_0f_{Q}=C[f_q,f_g,f_{Q}].
\end{equation}
We will consider only two-body collisions where the collision integral $C[f_q, f_g, f_Q] (p)$ can be expressed by the following relation:
\begin{align}\label{int_finale}
\begin{split}
&C[f]=\frac{1}{2E_p}\int\frac{d^3\textbf{q}}{2E_q (2\pi)^3}\int\frac{d^3\textbf{q}'}{2E_{q'}(2\pi)^3}\int\frac{d^3\textbf{p}'}{2E_{p'}(2\pi)^3}
\\&\cdot\frac{1}{d_Q}
\sum_{g,q,\bar q}|{\cal M}(g(q,\bar q) c\rightarrow  g(q,\bar q) c))|^2 
\\&\cdot (2\pi)^4\delta^4(p+q-p'-q')
[f_Q(\textbf{p}')\hat{f}(\textbf{q}')-f_Q(\textbf{p})\hat{f}(\textbf{q})]
\end{split}
\end{align}  
where \textbf{p} (\textbf{q}) and \textbf{p$\prime$} (\textbf{q$\prime$}) represent respectively the initial and final momentum of heavy quark (plasma particle) and $|{\cal M_Q}|^2$ is the squared modulus of scattering matrix of the process.
In order to solve the collision integral it is necessary to evaluate
the scattering matrix $|{\cal M_Q}|^2$. In our calculations the HQs interact with the medium by
mean of two-body collisions regulated by the scattering matrix of the processes $g + Q \to g + Q$ and $q(\bar q) + Q \to q(\bar q) + Q$. 

A successful way to treat non-perturbative effects in heavy-quark scattering is given by Quasi-Particle approach (QPM), in which the interaction is encoded in the quasi-particle masses that behave like massive constituents of free gas plus a background field interaction given by a temperature dependent bag constant, for details see ref.~\cite{Plumari:2011mk}. The main feature of QPM approach is that the resulting coupling is significantly stronger than the one coming from pQCD running coupling, particularly at $T \rightarrow T_c$. It has been shown that QPM can reproduce the lattice QCD Equation of State: pressure, energy density and interaction measure $T^\mu_\mu=\epsilon-3 P$.
The relations of the masses of light quarks and gluons to the coupling and temperature are calculated in a perturbative approach:

\begin{eqnarray}\label{masse_QPM}
&&m_g^2=\frac{1}{6}g(T)^2\left[\left(N_c+\frac{1}{2}N_f\right)T^2+\frac{N_c}{2\pi^2}\sum_{q}\mu_q^2\right],\nonumber \\
&&m_{u,d,s}^2=\frac{N_c^2-1}{8N_c}g(T)^2\left[ T^2+\frac{\mu^2_{u,d}}{\pi^2}\right]
\end{eqnarray}
where $N_f$ and $N_c$ are respectively the number of flavours and colours, $\mu_q$ is the chemical potential of the q flavour that in our calculation is neglected. Even if the formal relation is perturbative-like, the $g(T)$ is obtained by a fit to the energy density of lattice QCD (lQCD) and it is expressed by:
\begin{equation}
g^2(T)=\frac{48\pi^2}{[(11N_c-2N_f)ln[\lambda(\frac{T}{T_c}-\frac{T_s}{T_c})]]^2}
\end{equation}
where $\lambda=2.6$ and $T_s/T_c=0.57$, with $T_c=155 MeV$. We obtain a non perturbative behaviour of the coupling especially for $T\rightarrow T_c$.

The evaluation of the scattering matrix has been performed considering
the leading-order diagrams. In this approach the effective coupling 
$g(T)$ leads to effective vertices and a dressed massive gluon
propagator for $g + Q \to g + Q$ and massive
quark propagator for $q(\bar q)+ Q\to q (\bar q)+Q $ scatterings.
The detail of the calculations can be found in Ref. \cite{Berrehrah:2013mu}.

\subsection{On-shell Transport Coefficients and Fokker-Planch equation}
We give a brief description of the derivation of HQ transport coefficients.
We can also express the collision integral in relation to the rate of collisions $\omega(\textbf{p},\textbf{k})$ between HQ and light bulk particles, where \textbf{k} is the transferred momentum during the collision:
\begin{equation}\label{integrale_coll}
C[f]=\int d^3\textbf{k} [\omega (\textbf{p}+\textbf{k},\textbf{k})f(\textbf{p}+\textbf{k})-\omega(\textbf{p},\textbf{k})f(\textbf{p})].
\end{equation}
The rate of collision $\omega(\textbf{p},\textbf{k})$ is given by:
\begin{equation}\label{rate}
\omega(\textbf{p},\textbf{k})=d_{QGP} \int \frac{d^3\textbf{q}}{(2\pi)^3} \hat f(\textbf{q})v_{q,p}
\frac{d\sigma_{p,q\rightarrow p-k,q+k}}{d\Omega}.
\end{equation}
In this relation $d_{QGP}$ defines degree of freedom of particle in collision with heavy quark, $\hat f(\textbf{q})$ is the time and space independent distribution function of particle in the plasma of momentum $\textbf{q}$, $v_{q,p}$ defines the relative velocity and $\sigma_{p,q\rightarrow p-k,q+k}$ is the differential cross section of the scattering process.

Differential cross section can be expressed by the following relation:
\begin{equation}\begin{split}
\frac{d\sigma_{p,q\rightarrow p-k,q+k}}{d\Omega}=\frac{1}{(2\pi)^6}\frac{1}{v_{p,q}}\frac{1}{2E_q}\frac{1}{2E_p
}\frac{1}{d_Q d_{QGP}}
\sum|{\cal M_Q}|^2\\\times\frac{1}{2E_{q+k}}\frac{1}{2E_{p-k}}(2\pi)^4\delta(E_p+E_q-E_{p-k}-E_{q+k})
\end{split}
\end{equation}
with the $\sum$ intended over all the elastic scattering channels $g+Q \rightarrow g+Q$ and $q(\bar q)+ Q\rightarrow q (\bar q)+Q $.
The non-linear integro-differential Boltzmann equation cannot be easily solved and a way to simplify the calculation is to employ the Landau approximation leading to a relativistic Fokker-Planck equation with momentum
dependent transport equation.
This assumption is physical motivated by the suggestion that during collision the transfer momentum $\textbf{k}$ is small and we can operate an expansion of the integral:
\begin{align}
\begin{split}
f(\textbf{p}+\textbf{k})\omega(\textbf{p}+\textbf{k},\textbf{k})=&f(\textbf{p})\omega(\textbf{p},\textbf{k})+\textbf{k}\frac{\partial}{\partial \textbf{p}}(\omega f)\\&\times+\frac{1}{2}k_ik_j\frac{\partial^2}{\partial p_i \partial p_j}(\omega f)+...
\end{split}
\end{align} 
Defining the following quantities:
\begin{equation}\begin{split}\label{coeff}
A_i(\textbf{p},T)&=\int d^3kk_i\omega(\textbf{p},\textbf{k})\\
B_{i,j}(\textbf{p},T)&=\frac{1}{2}\int d^3kk_ik_j\omega(\textbf{p},\textbf{k})
\end{split}
\end{equation}
the collision integral $C[f]$ in Eq.\ref{int_finale} becomes:
\begin{eqnarray}\label{fokker-planck}
\frac{df(\textbf{p})}{dt}&=&\frac{\partial}{\partial p_i}\left[A_i(\textbf{p},T)f(\textbf{p})+\frac{\partial}{\partial p_j}[B_{i,j}(\textbf{p},T)f(\textbf{p})]\right]. \nonumber \\
\end{eqnarray}

The Eq.\ref{fokker-planck} is the Fokker-Planck equation and the quantities defined by the Eq.\ref{coeff} are the drag and diffusion coefficients of the propagation of HQ in the thermal bath at temperature T.
If we consider an isotropic medium, we can express the drag and diffusion coefficients by the following relations:
\begin{equation}\begin{split}
A_i(\textbf{p},T)&=A(p,T)p_i\\
B_ {i,j}(\textbf{p},T)&=B_L(p,T)P_{i,j}^{||}(\textbf{p})+B_T(p,T)P_{i,j}^{\perp}(\textbf{p}).
\end{split}
\end{equation}
The diffusion coefficient is expressed by a longitudinal $B_L$ and a transversal $B_T$ component respect to the HQ momentum where $P_{i,j}^{||}(\textbf{p})=p_ip_j/\textbf{p}^2$ and $P_{i,j}^{\perp}(\textbf{p})=\delta_{i,j}-(p_ip_j/\textbf{p}^2)$ are the projection operators on the longitudinal and transverse momentum components.
Using the definition in Eq.\ref{coeff} we get the following expression for $A_i$:
\begin{eqnarray}\label{drag}
  & & A_i(\textbf{p},T)= \nonumber \\
  &=&\frac{1}{2E_p}\int\frac{d^3\textbf{q}}{2E_q (2\pi)^3}\int\frac{d^3\textbf{q}'}{2E_{q'}(2\pi)^3}\int\frac{d^3\textbf{p}'}{2E_{p'}(2\pi)^3}\nonumber \\
  &\times&\frac{1}{d_Q}\times \sum|{\cal M_Q}|^2(2\pi)^4\delta^4(p+q-p'-q') \nonumber \\
  &\times&\hat f(q)[(p-p')_i]\equiv \left\langle \left \langle(p-p')_i\right \rangle \right\rangle.
\end{eqnarray}
while for $B_{i,j}$:
\begin{equation}\label{diff}
B_{i,j}(\textbf{p},T)=\frac{1}{2}\left\langle \left \langle(p-p')_i(p'-p)_j\right \rangle \right\rangle.
\end{equation}
Finally, drag, transverse and longitudinal diffusion coefficients can be calculated as follows:
\begin{align}\begin{split}\label{B1_comp}
B_L(p,T)&=\frac{1}{2}\frac{p_ip_j}{p^2}B_{i,j}=\\
&=\frac{1}{2}[\left\langle \left \langle\textbf{p}'\cdot \textbf{p}\right \rangle \right \rangle p^2-2\left\langle \left \langle\textbf{p}'\cdot \textbf{p}\right \rangle \right \rangle+p^2\left\langle \left \langle\textbf{1}\right \rangle \right \rangle]
\end{split}
\end{align}
\begin{equation}\begin{split}\label{B0_comp}
B_T(p,T)&=\frac{1}{2}\left[\delta_{i,j}-\frac{p_ip_j}{p^2}\right]B_{i,j}=\\&=
\frac{1}{4}[\left\langle \left \langle p'^2\right \rangle \right \rangle-\left \langle \left \langle(\textbf{p}\cdot \textbf{p}')^2\right \rangle \right \rangle/p^2].
\end{split}
\end{equation}
and for drag coefficient:
\begin{equation}\begin{split}\label{drag_comp}
A(p,T)&=p_iA_i/p^2=\\&=\left\langle \left \langle\textbf{1}\right \rangle \right \rangle-\left\langle \left \langle\textbf{p}'\cdot\textbf{p}\right \rangle \right \rangle/p^2.
\end{split}
\end{equation}
We recall that the standard approach to evaluate the quantities in Eq.\ref{drag} and Eq.\ref{diff} is to write the integral in the c.m. frame using the c.m. scattering angles and the momentum $\textbf{q}$ of the plasma particle:
\begin{align}\begin{split}\label{F_finale}
\left\langle \left \langle F(\textbf{p},\textbf{p}\prime,T)\right \rangle \right\rangle =\frac{1}{(2\pi)^3}\int_{0}^{\infty} d\textbf{q}\textbf{q}^2\int_{-1}^{+1} dcos\alpha\\\times \int_{t_{min}}^{t_{max}} dt v_{rel} \frac{d\sigma}{dt} 
\hat f(\textbf{q}) \int_{0}^{2\pi} d\phi_{cm} F(\textbf{p},\textbf{p}\prime,T)
\end{split}
\end{align} 
where $\alpha$ is the polar angle of $\textbf{q}$ and the Mandelstam variable $t$ is expressed in terms of momentum $\hat{\textbf{p}}$ of heavy quark in the c.m. scattering system by $t=(p-p')^2=-2|\hat{\textbf{p}}|^2(1-cos\theta_{cm})$.
Finally, the differential cross section takes the form:
\begin{equation}
\frac{d\sigma}{dt}=\frac{1}{16\pi}\frac{1}{[(s-M_Q^2-m^2)^2-4M_Q^2m^2]}\frac{1}{d_Q}\sum|{\cal M_Q}|^2.
\end{equation}

\subsection{Off-shell Transport coefficients}
In order to have a more accurate description, the propagation of heavy quark can be also treated taking into account 
off-shell effects due to collisions with quasi-particle in the plasma. The collision integral, that in the on-shell case is expressed by Eq.\ref{int_finale}, in the off-shell case can be written as:

\begin{eqnarray}\label{int_finale_off}
  C[f]&=&\int dm_i A(m_i)\int dm_fA(m_f) \nonumber \\
  &\times&\frac{1}{2E_p}\int\frac{d^3\textbf{q}}{2E_q (2\pi)^3}\int\frac{d^3\textbf{q}'}{2E_{q'}(2\pi)^3}\int\frac{d^3\textbf{p}'}{2E_{p'}(2\pi)^3}\nonumber\\
  &\times&\frac{1}{\gamma_Q}\sum|{\cal M_Q}|^2(2\pi)^4\delta^4(p+q-p'-q')\nonumber\\
  &\times&[f(\textbf{p}')\hat{f}(\textbf{q}',m_f)-f(\textbf{p})\hat{f}(\textbf{q},m_i)]
\end{eqnarray}  

In particular, we want to investigate how the off-shell quantum effects modify the evolution of charm quark respect 
to the on-shell case that as a first approximation is commonly used to study the propagation of these particles 
in the bulk of light quarks and gluons.

The dynamical quasi-particle model (DQPM) describes QCD properties in terms of the
single-particle Green's functions which leads to the 
description of QGP in terms of strongly interacting
massive effective quasi-particles with broad spectral functions \cite{Cassing:2009vt}.
In this approach the parton masses and widths are determined by fitting 
the quasi-particle entropy density to the lQCD entropy density reproducing the QCD
equation of state extracted from lattice QCD calculations \cite{Bratkovskaya:2011wp}.
The aim of this study is an evaluation of the off-shell effects due to plasma quasi-particles.
In the DQPM approach in Ref.\cite{Berrehrah:2013mu}, partons are dressed by non perturbative
spectral function $A(q^0)$ which associates a spectrum of energies to a particle of momentum $\textbf{q}$.
The ansatz used to model a nonzero width is obtained 
by replacing the free spectral function by a Lorentzian form \cite{Berrehrah:2013mu}.
As shown in Ref.\cite{Berrehrah:2013mu} the Lorentzian form has a peak at small values of $p/T$ 
at the pole mass of the charm quarks and for it a non-relativistic approximation 
is a good approximation. 

In this work, we are interested in a non-relativistic approximation of partonic spectral
function in which at small momenta $q^0\approx m$, in this way $A(q^0)$ is parametrized by a Breit-Wigner
function $A^{BW}(m_i)$ \cite{Berrehrah:2013mu, Bratkovskaya}.
The width of the partons in the perturbative limit are given by  
$\gamma \approx g^{2} T \ln{g^{-1}}$ where the physical process 
contributing to the functional form of the widths are elastic 
scattering like $gg \to gg$, $gq \to gq$ and $qq(\bar{q}) \to qq(\bar{q})$
are included.
The functional forms of bulk particles widths $\gamma_g$ and $\gamma_q$
associated to spectral function for $\mu_q=0$ are given by:
\begin{equation}\begin{split}\label{largh_QPM}
\gamma_g(T)&=\frac{1}{3}N_C\frac{g^2(T/T_C)}{8\pi} T\, ln\left[\frac{2c}{g^2(T/T_C)}+1\right]\\
\gamma_q(T)&=\frac{1}{3}\frac{N_C^2-1}{2N_C}\frac{g^2(T/T_C)}{8\pi} T\, ln\left[\frac{2c}{g^2(T/T_C)}+1\right]
\end{split}
\end{equation}
Fitting the entropy density on the lQCD data, the constant $c$ is fixed to $c=14.4$.
The spectral function associated to light quark and gluon in the plasma are expressed by:
\begin{equation}\label{Breit-Wigner}
A_i^{BW}(m_i)=\frac{2}{\pi}\frac{m_i^2\gamma_i^\ast}{(m_i^2-M_i^2)^2+(m_i\gamma_i^\ast)^2} 
\end{equation}
where $A_i^{BW}$ fulfills the normalization
\begin{equation*}
\int_{0}^{\infty}dm_iA_i(m_i,T)=1.
\end{equation*}

In Eq. \ref{Breit-Wigner}, $M_i$ is the pole mass of gluon and light quark defined in Eq.\ref{masse_QPM} and $\gamma_i^*$ is the width associated to each particle mass. Such widths are related to $\gamma_i$ calculated in DQPM approach by the relation $2q^0_i\gamma_i=m_i\gamma^*_i$ \cite{Berrehrah:2013mu}. Since we are taking into account 
a regime where $\gamma<M_i$, the $\gamma_i^\ast$ of Eq. \ref{Breit-Wigner} can be written  $\gamma_i^\ast \approx 2\gamma_i$. 
The off-shell dynamics implies that the values of partonic masses can be different before and after scattering process, i.e. $m_i\neq m_f$, differently from on-shell case in which $m_i = m_f$, where Breit-Wigner function becomes a delta function centered at Pole mass $M_i$.
\begin{figure}[ht]
  \begin{center}
    \includegraphics[width=\linewidth]{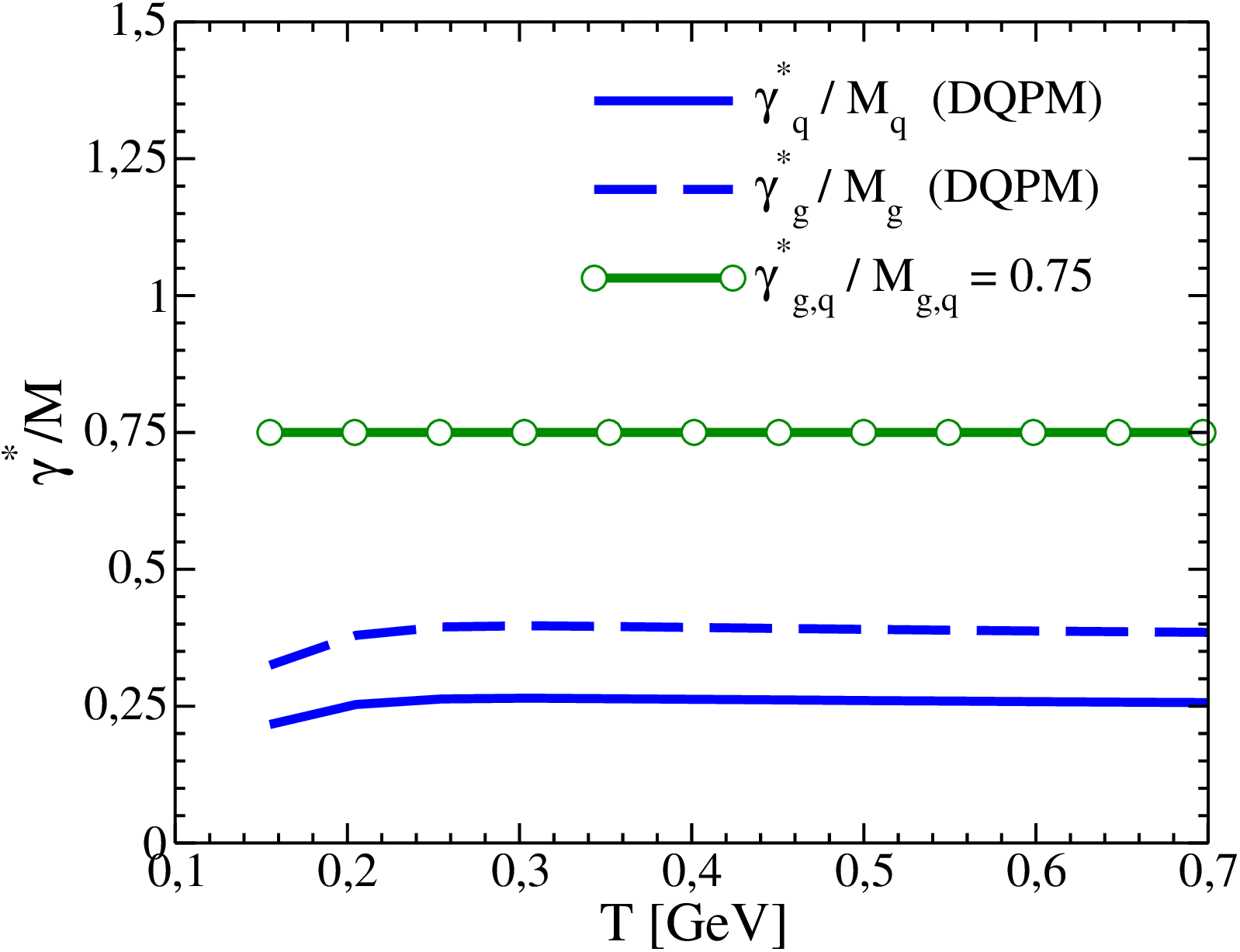}
    \caption{Ratio between widths calculated in DQPM approach and the Pole Mass $M_i$ as function of temperature for quarks (blue solid line) and gluons (blue dashed line). The green solid line is the correponding case with constant ratio fixed to $\gamma^*_i/M_i=0.75$.}\label{gamma_m}
  \end{center}
\end{figure}
In Fig. \ref{gamma_m}, ratio between $\gamma^*_i$ and Pole mass $M_i$ are shown for gluon and light quark. \\

If we consider the off-shell quantum effects for the plasma particle, the quantity of Eq.\ref{F_finale} can be written:
\begin{align}\begin{split}\label{F_finale_os}
\left\langle \left\langle F(m_i,m_f,\textbf{p},\textbf{p}\prime,T)\right\rangle \right\rangle  =\int dm_i A(m_i)\int dm_fA(m_f)&\\\times 
\frac{1}{(2\pi)^3}\int_{0}^{\infty} d\textbf{q}\textbf{q}^2\int_{-1}^{+1} dcos\alpha\int_{t_{min}}^{t_{max}} dt   v_{rel}&\\\times \frac{d\sigma}{dt} 
\hat f(m_i,\textbf{q}) \int_{0}^{2\pi} d\phi_{cm} F(m_i,m_f,\textbf{p},\textbf{p}\prime,T)
\end{split}
\end{align} 

where $m_i$ and $m_f$ are respectively the initial and final mass of partons respectively.
In this case the Mandelstam variable
$t=(p-p')=2M_Q^2-2\hat{E}_p\hat{E}_{p'}+2|\hat{\textbf{p}}||\hat{\textbf{p}}'|cos\theta_{cm}$.

\section{Results for transport coefficients: on-shell and off-shell}\label{Section:3}
In the following we will compare the results coming from the on-shell expression in Eq.\ref{F_finale} with the one in Eq.\ref{F_finale_os} that include the off-shell effects.

Before systematically study and compare the transport coefficients between the two different approaches presented 
in the previous sections we describe the common features in the following calculations.
The number of thermal quark flavors is set to $n_f = 3$, the medium temperature is kept fixed. 
A Boltzmann distribution is used for the thermal light flavor quark and gluon distribution.
The charm quark mass is fixed to $M_c=1.3 \, GeV$.
To regulate the collinear divergence of the t-channel in the scattering matrix,
the following replacement is performed $1/t \to 1/(t - \mu_D^2 )$.
Where we have set the Debye screening mass to $m_D=\sqrt{4\pi\alpha_s(T)}T=g(T)T$.
In the off-shell case this replacement takes the form $1/t \to 1/(t - \mu_D^2+i2\gamma_g(p^0_f-p^0_i) )$ where $p^0$ is the energy of charm quark \cite{Berrehrah:2013mu}.

\begin{figure}[ht]
\begin{center}
		\includegraphics[width=\linewidth]{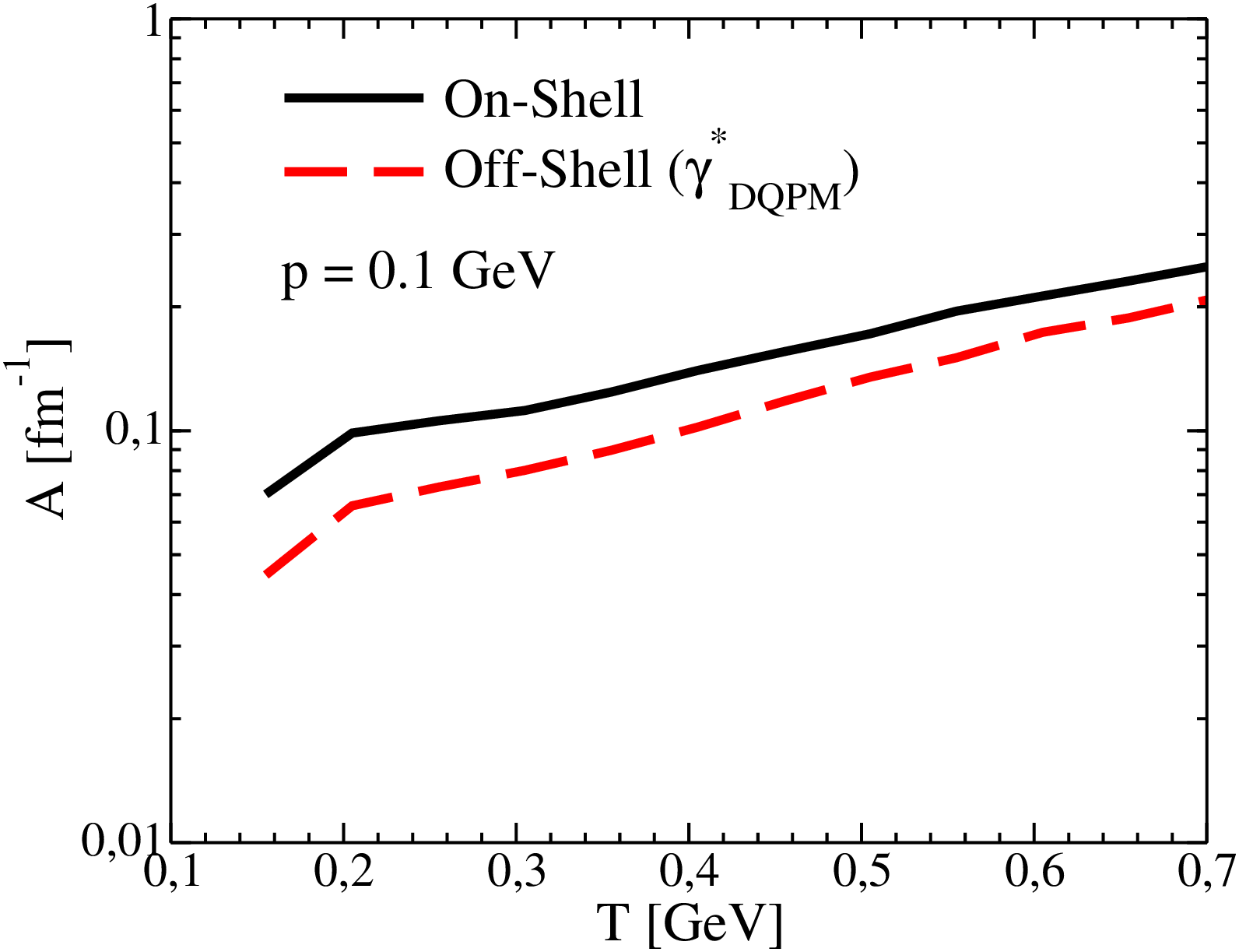}
	\caption{Drag coefficient ($A(T)$) as a function of the medium temperature at fixed HQ momentum $p=0.1 \, GeV$ for on-shell approach (black solid line) and off-shell approach (red dashed line).}\label{Fig:drag_T}
\end{center}
\end{figure}

\begin{figure}[ht]
\begin{center}
		\includegraphics[width=\linewidth]{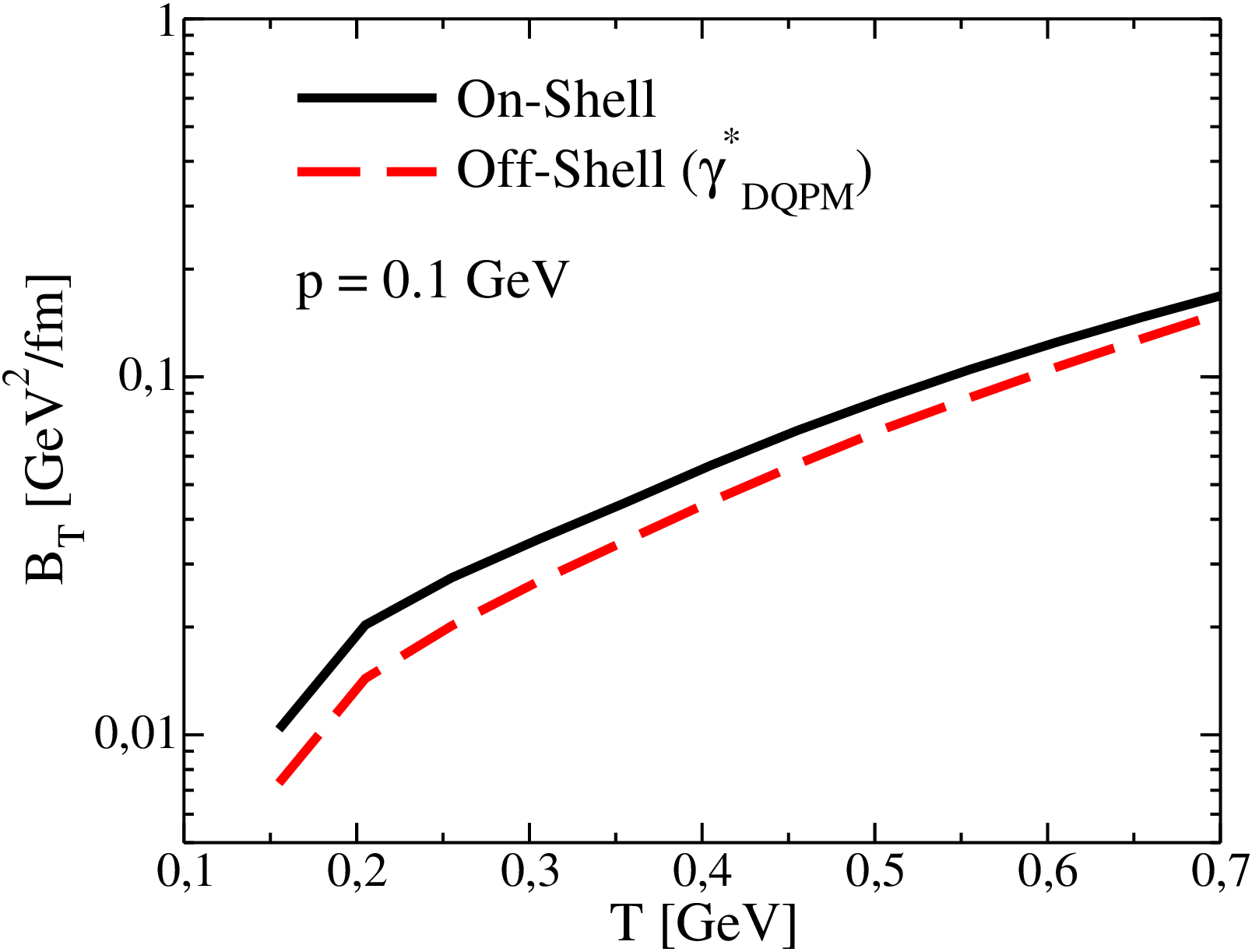}
	        \caption{Diffusion coefficient $B_T(T)$ as a function of the medium temperature at fixed HQ momentum $p=0.1 \, GeV$ for on-shell approach (black solid line) and off-shell approach (red dashed line).}\label{Fig:BT_T}
\end{center}
\end{figure}

In Fig.\ref{Fig:drag_T} and Fig.\ref{Fig:BT_T} we compare the transport coefficients as a function of the medium temperature with a fixed HQ momentum of $p = 0.1$ $GeV/c$ for the two different approaches studied in this paper. Solid lines refers to on-shell calculations while red dashed lines for off-shell calculations.
In Fig.\ref{Fig:drag_T} we show the results for the drag coefficient $A$. 
If quantum off-shell effects for bulk are considered the drag coefficients decrease of about $30 \%$ in the temperature regime of $T\sim1-2T_c$.
Similar conclusion we get also for the diffusion coefficient $B_T$ as show in Fig.\ref{Fig:BT_T}, where in this case we observe a reduction of about $35$ $\%$ in the same range of temperature. Both difference decrease at increasing temperature.
We clarify that here we are keeping the couplig of quark and gluons to be 
the same in the on-shell and off-shell case to see the main direct effect of the inclusion of a finite widths
for the quasi-particles. Of course another approach could be to upscale $g(T)$ to have the same energy density
in both on-shell and off-shell case. Being the change in energy density of about $10-15\%$ this case
corresponds to a change of $g(T)$ by only few percent.

\begin{figure}[ht]
	\begin{center}
	\includegraphics[width=\linewidth]{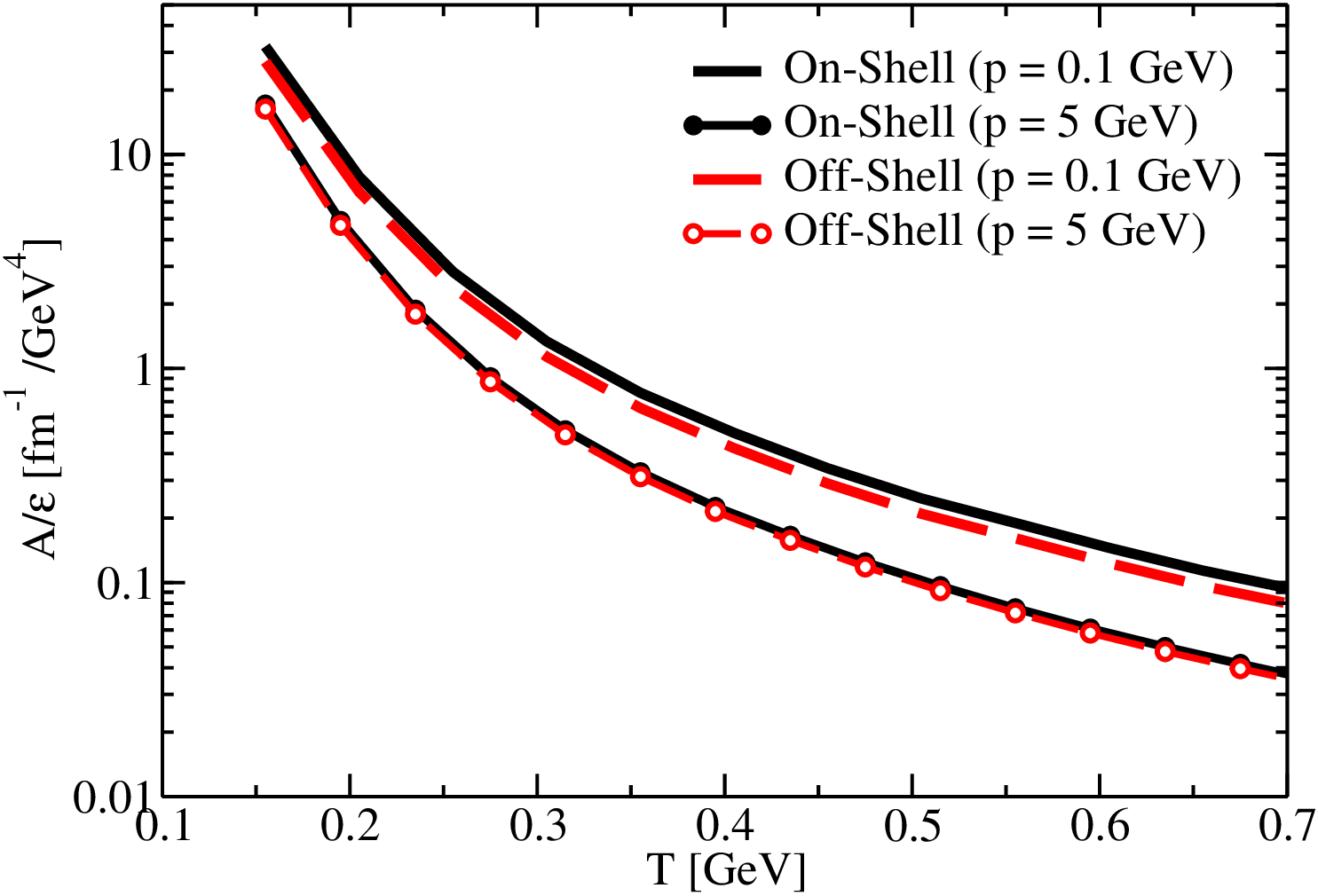}
	\caption{Ratio between drag coefficient ($A$) and energy density ($\epsilon$) as function of temperature for two different HQ momentum $p=0.1 \, GeV$ and $p=5 \, GeV$. Black solid and red dashed line respectively for  on-shell and off-shell for $p=0.1 \, GeV$ while full black circles and open red circles respectively for on-shell and off-shell for $p=5 \, GeV$}\label{Fig:drag_eps_T}
\end{center}
\end{figure}

\begin{figure}[ht]
	\begin{center}
		\includegraphics[width=\linewidth]{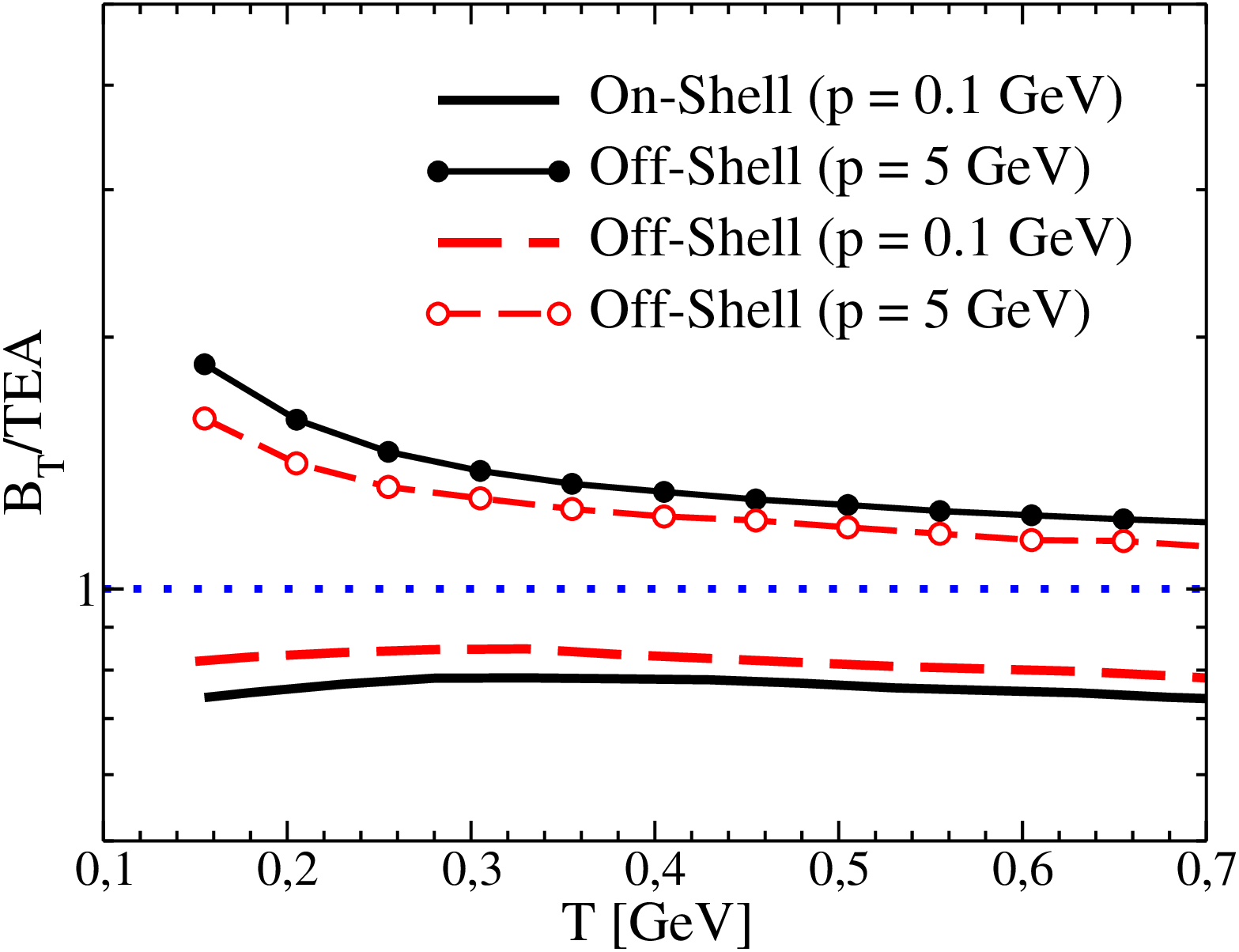}
		\caption{$B_T/TEA$ as a function of temperature for both on-shell and off-shell approaches and $p=0.1 \, GeV$ and $p=5 \, GeV$. Same legend as in Fig.\ref{Fig:drag_eps_T}.}\label{Fig:FDT}
	\end{center}
\end{figure}

We have checked if the decrease in the drag can be a mere effect of the decrease of the equilibrium energy density $\epsilon$. 
Therefore we have calculated the $A/\epsilon$ ratio as shown in Fig.\ref{Fig:drag_eps_T} for different values of temperature.
We can see that when we divide drag coefficient obtained in on-shell and off-shell mode by the respective values of the 
energy density of the bulk system, the decrease of coefficient in off-shell case is completely re-absorbed 
for intermediate and high momentum. 
The scaling with $\epsilon$ is only partially fulfilled in the limit $p\rightarrow 0$ and we see that a difference 
between on-shell and off-shell mode of about $10 \%$ remains even when the comparison is done
renormalizing at the same energy density. 
This allows to draw a first conclusion about the fact that there is an impact 
of off-shell effects at low $p$, but a sizeable part can be traced-back to a change of the energy density
when simply increasing increasing the width. 
Furthermore already at intermediate momenta $p$ the transport coefficients become the same once
renormalized to the energy density,
as we can see in Fig. \ref{Fig:drag_eps_T} comparing open and filled circles.
This is true at least when the values of the widths are relatively small as in DQPM approach.
In this context, we want also to check the violation of fluctuation-dissipation theorem (FDT) \cite{Das:2013kea} 
for on-shell and off-shell case. The validity of this relation can be verified evaluating the ratio between diffusion 
coefficient $B_T$, obtained by scattering matrix ${\cal M_Q}$ with the value of $B_T$ predicted by fluctuation-dissipation relation. 
In order to fulfill the FDT this ratio should be equal to 1. In general, when we calculate transport coefficient 
with scattering matrix one obtains however a significant deviation \cite{Das:2013kea}.
In Fig.\ref{Fig:FDT} it is shown the ratio between $B_T$ and $TEA$ where $E=\sqrt{p^2+M^2}$ for the two cases discussed in this paper for on-shell and off-shell partons. We observe that the FDT is better verified when off-shell bulk is taken into account where 
we obtain an improvement with respect to on-shell case of about $10 \%$. Moreover we observe that the FDT is better verified 
at higher temperature where we obtain a deviation lower than $15 \%$.
In the results shown in the previous figures we have considered the widths $\gamma^*_i$ given by the DQPM approach\cite{Berrehrah:2013mu}. As shown in Fig.\ref{gamma_m}, such widths (i.e. $\gamma^*_{q}\approx 260$ MeV, $\gamma^*_{q}\approx 110$ MeV at $T=200$ MeV) are significant smaller than the quasi-particle masses.

In order to explore also the impact of quantum off-shell effects on the transport coefficients we artificially increase 
the widths considering $\gamma^*$ about 2-3 times larger than those of DQPM approach 
for quarks, i.e. $\gamma^*/M=0.75$ 
for both quarks and gluons (i.e. $\gamma^*_{g}\approx 520$ MeV, $\gamma^*_{q}\approx 330$ MeV at $T=200$ MeV) 
as shown by green line in Fig.\ref{gamma_m}. 
We consider  larger widths with respect to the DQPM, because they are conceivable 
in other approaches especially considering values of the shear viscosity
to entropy density  ratio, $\eta/s$, is about 0.1, while for DQPM it is stays in the range $\eta/s \sim 0.2 - 0.3$
for $T \sim T_c$.

\begin{figure}[ht]
	\begin{center}
		\includegraphics[width=\linewidth]{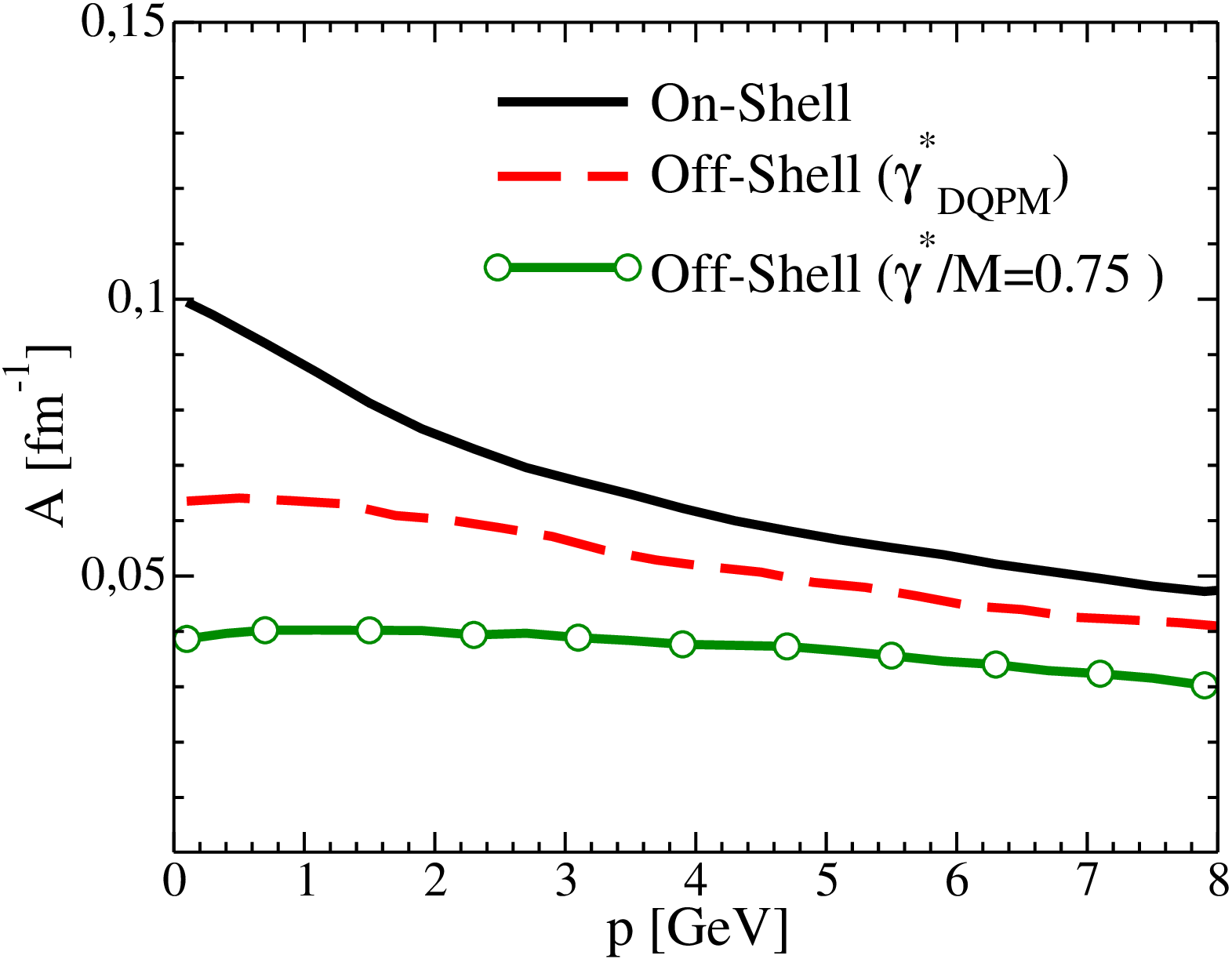}
		\caption{Drag coefficient ($A$) as a function of momentum for fixed medium temperature at $T=0.2 \, GeV$ for on-shell approach (black solid line) and off-shell approach (red dashed line). The open green circles refers to the case with larger widths with fixed ratio $\gamma^*_i/M_i=0.75$.}\label{drag200}
	\end{center}
\end{figure}

\begin{figure}[ht]
	\begin{center}
		\includegraphics[width=\linewidth]{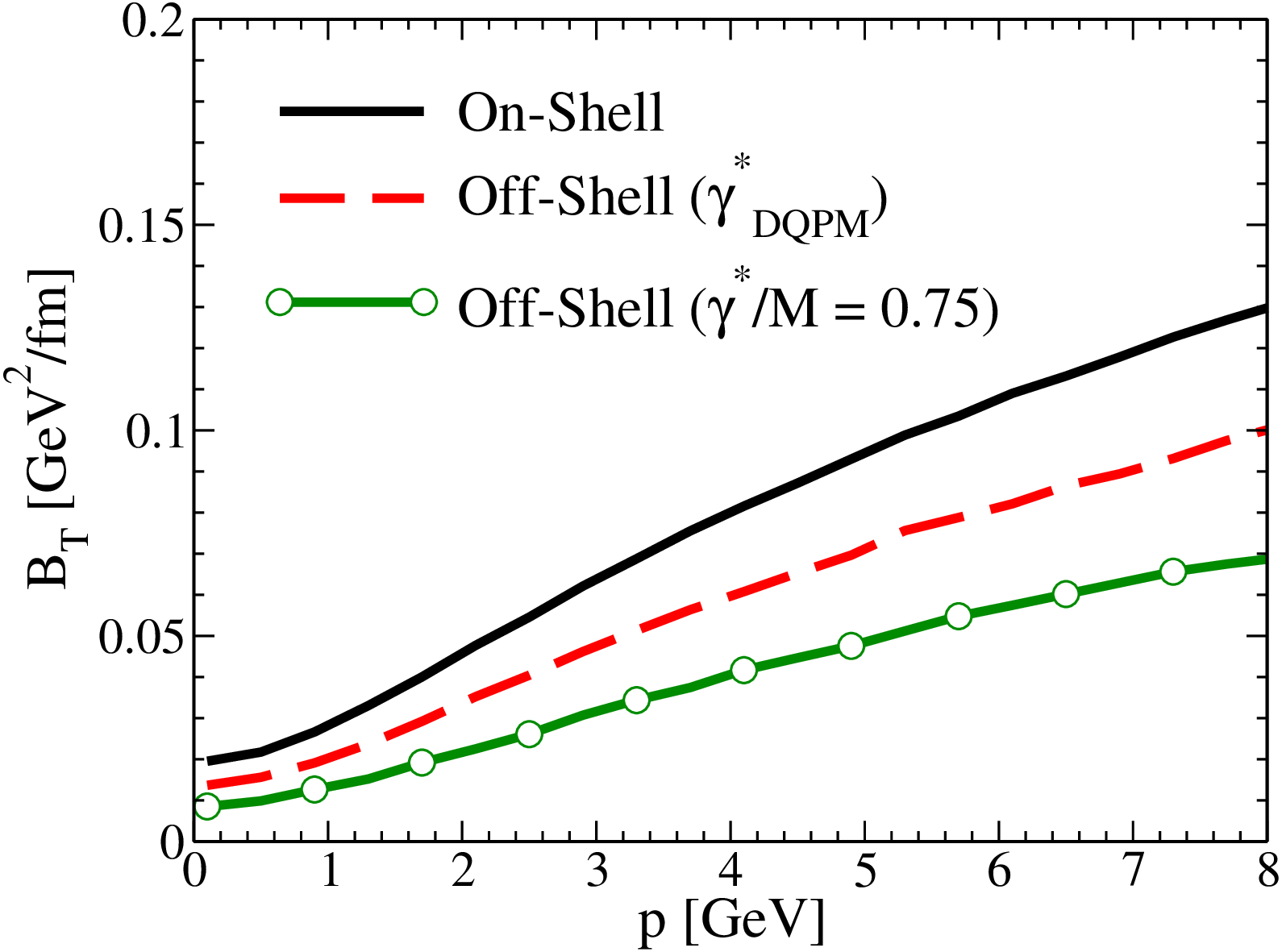}
	        \caption{Diffusion coefficient ($B_T$) as a function of momentum and for fixed medium temperature at $T=0.2 \, GeV$ for on-shell approach (black solid line) and off-shell approach (red dashed line). The open green circles refers to the case with larger widths with fixed ratio $\gamma^*_i/M_i=0.75$.}\label{BT200}
\end{center}
\end{figure}

\begin{figure}[ht]
	\begin{center}
		\includegraphics[width=\linewidth]{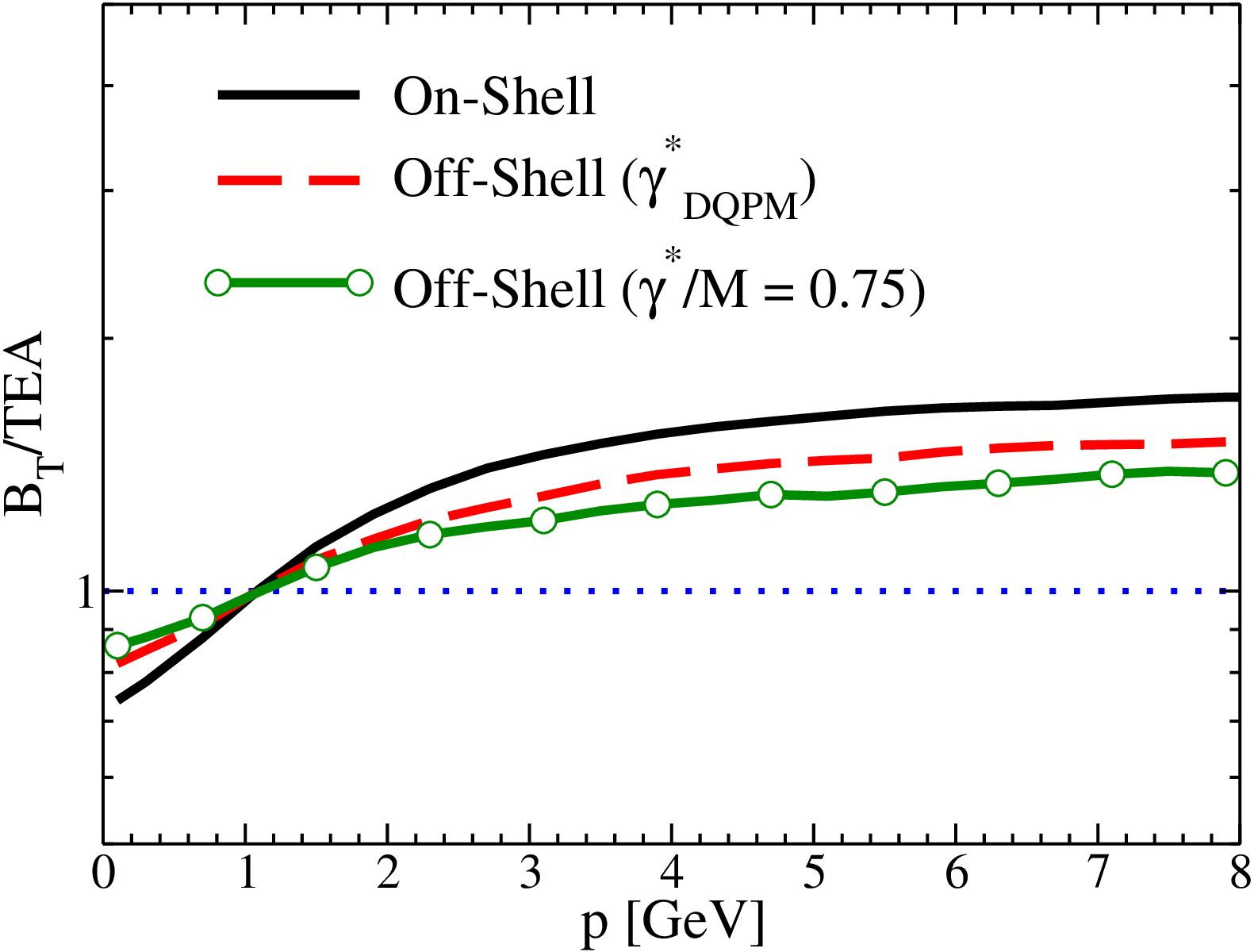}
	        \caption{$B_T/TEA$ as a function of charm momentum and fixed temperature $T=0.2 \, GeV$ for both on-shell and off-shell approaches. Same legend as in Fig.\ref{drag200}.}\label{FDT200}
\end{center}
\end{figure}

\begin{figure}[ht]
	\begin{center}
		\includegraphics[width=\linewidth]{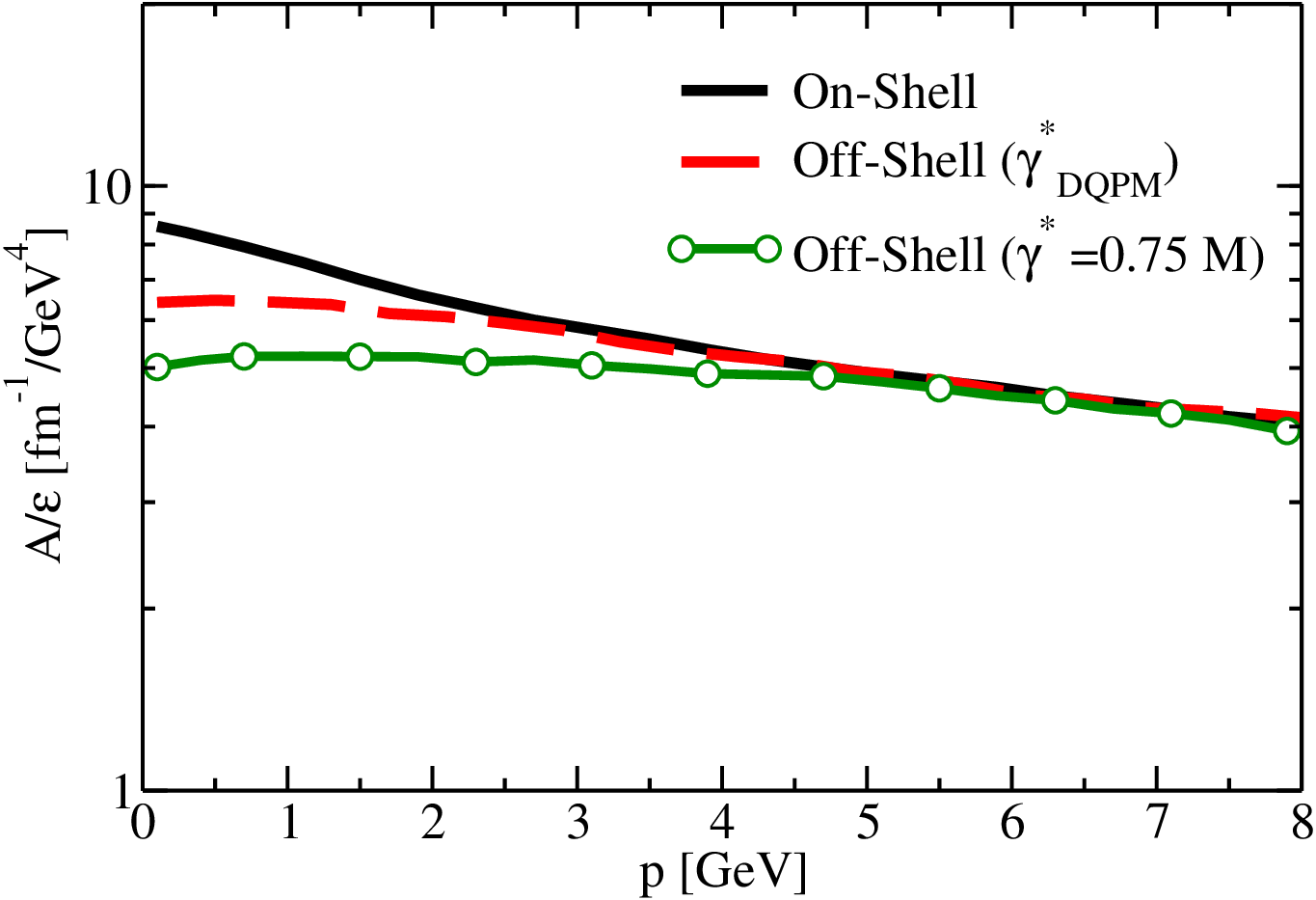}
	\caption{$A(p)/\epsilon$ ratio as function of charm momentum at $T=0.2$ GeV for on shell (black solid line)
	and off-shell $\gamma^*_{DQPM}$ (red dashed line) and $\gamma^*=0.75 M$ (open green circles). }\label{Fig:drag_eps}
\end{center}
\end{figure}

In Fig.\ref{drag200} and Fig.\ref{BT200} we have shown the HQ transport coefficients, respectively drag $A$ and diffusion $B_T$, as function of charm momentum at fixed $T=0.2 \, GeV$ including now the $\gamma^*/M=0.75$ case. 
If we can see that considering bigger widths for Breit-Wigner distribution with respect to DQPM one, 
there is a decrease of the transport coefficients. 
Furthermore, a limited improvement for FDT validity is observed for the case of larger widths where the FDT is satisfied 
within $10$ $\%$, as shown by open green circles in Fig.\ref{FDT200}.
Finally, in Fig.\ref{Fig:drag_eps} it is shown the $A(p)/\epsilon$ ratio as a function of the charm quark momentum and for 
temperature $T=0.2 \, \rm GeV$. Comparing black solid line with red dashed line, we observe a scaling between on-shell and off-shell calculation for 
$p \ge 2-3 \, \rm GeV$ and breaking a lower momentum. 
This suggests that for the widths used in DQPM the difference in the drag coefficient in off-shell case is completely re-adsorbed for high momentum of charm. Furthermore, if we increase the widths as shown by open green circles we get that the drag coefficient 
shows a larger breaking of the scaling at least 
for $p\lesssim 2-3 \, \rm GeV$ that at $p \rightarrow 0$ is maximal and corresponds to a reduction of about a $40 \%$.

\section{Heavy quarks momentum evolution in the QGP: on- and off-shell Boltzmann and Langevin dynamics}
\label{Section:4} 

In this section we discuss about the time evolution of HQs within Boltzmann scattering with on-shell quarks and gluons and the extension of the Boltzmann collision integral to account for off-shell conditions by mean of the Breit-Wigner spectral functions as done for the transport coefficients, see Eq. \ref{int_finale_off}. We are interested in the evolution of the HQ distribution function $f_{Q}(x,p)$ in a thermal bulk described through QPM approach and we have considered a plasma in equilibrium in a box with constant temperature T.
In this study, the starting point to investigate the HQ evolution for both on-shell and off-shell approaches is the simplified form of the Boltzmann equation that is expressed in Eq.\ref{Boltz_simply}. We can write: 
\begin{equation}\label{bolt_vera}
\frac{\partial f_{Q}}{\partial t}=\frac{1}{E_{Q}}C[f_q,f_g,f_{Q}].
\end{equation}

Since in the previous equation the field gradients are discarded, it is valid for both on-shell 
and off-shell dynamics, with the last embedded in $C[f_q,f_g,f_{Q}]$ according to Eq. \ref{int_finale_off}.
After a time discretization the Boltzmann equation can be written as

\begin{equation}\label{Eq:discr}
f(t+\Delta t,p)=f(t,p)+\frac{\Delta t}{E_{Q}}C[f] + O(\Delta t^3).
\end{equation}

As for the transport coefficients, the numerical solution of Boltzmann equation is obtained by a code that implement a Monte-Carlo integration method for the full collision kernel described by Eq.s \ref{int_finale} and \ref{int_finale_off}. Different tests have been done in order to verify the convergency of the collision integrals both in on-shell and off-shell case. It is important to fix the number of MonteCarlo samples $N_s$, in particular for off-shell case, where we have two additional integrations over spectral function that give the weight of each initial and final mass of light partons in the bulk.
In this study, we have discretized the time and the HQ momentum \textit{p} in the propagation in order to calculate the evolution of phase-space distribution function of charm quarks.
We want that the integral over the distribution function is conserved. Therefore, we can write:
\begin{equation}
\frac{\partial N}{\partial t}=\int d^3p \frac{\partial f}{\partial t}=\int d^3p \frac{C[f]}{E_{Q}}\equiv \bar{C}
\end{equation}

where $N$ is the number of charms quarks. If the integral is not conserved, we can assume a variation $\Delta N$ according to $N(t)=N_0+\bar{C}\Delta t$ where $N_0$ is the initial number of charm quarks.

\begin{figure}[t]
	\begin{center}
		\includegraphics[width=\linewidth]{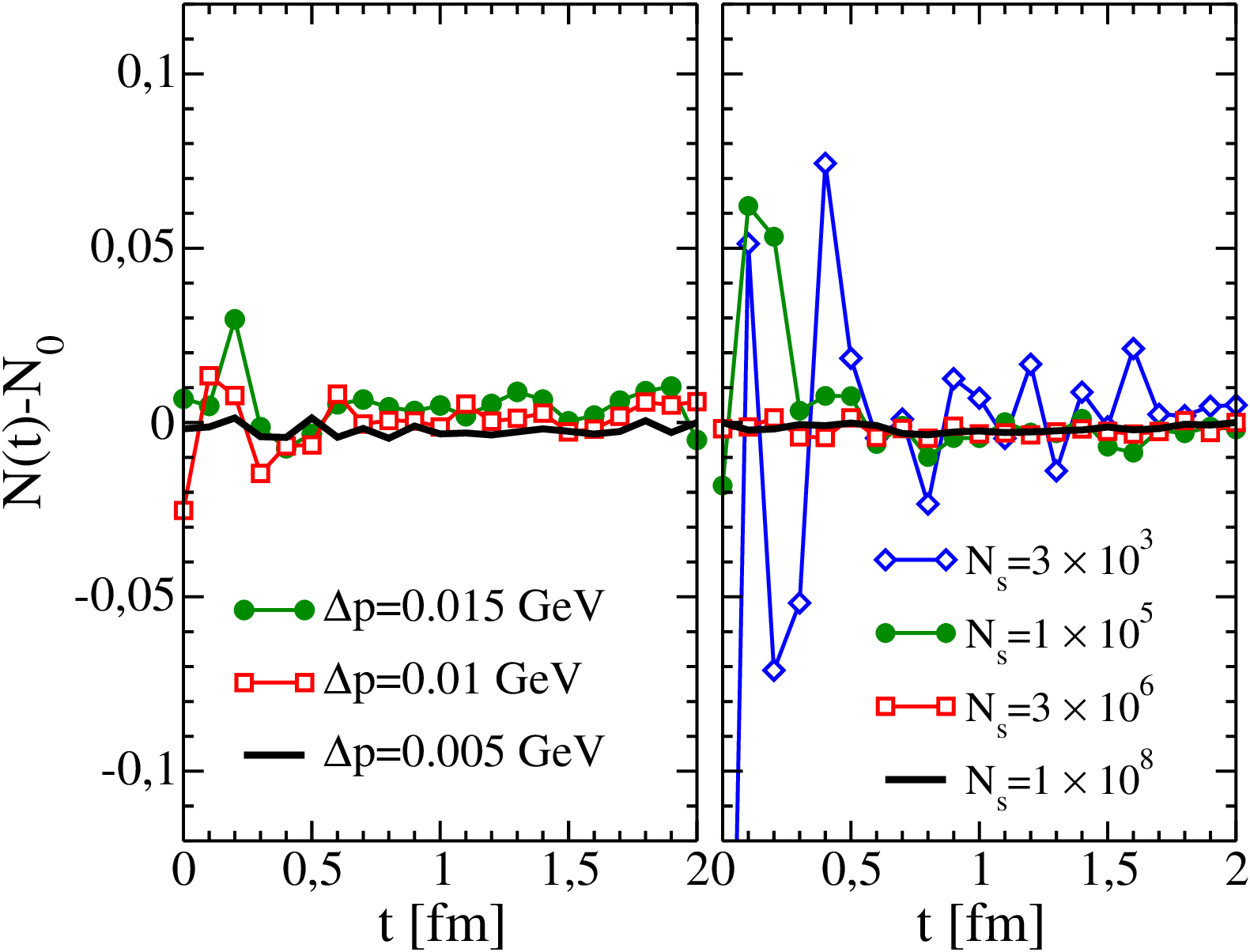}
	        \caption{$N(t)-N_0$ as a function of time. Left panel different lines are for different momentum discretization with a fixed number of MC sampling to $N_s=10^8$ while in the right panel the different lines refers to the different MC sampling used for $\Delta P= 0.005 GeV$.}\label{conv_dp}
\end{center}
\end{figure}
In Fig.\ref{conv_dp} it is shown an example of study of the convergence of the off-shell collision integral, similar results 
we get also for the on-shell case. In particular, we have studied the time evolution of $\Delta N$ for different momentum 
discretization $\Delta p$ (left panel) and number of samples $N_s$ (right panel) used for the Monte-Carlo calculation of the collision integral.
We have found that the most appropriate number of momentum discretization and Monte-Carlo samples is 
$\Delta p =5 \times 10^{-3} GeV$ and $N_s=10^8$. A similar study it has been performed for the time step $\Delta t$ we found that 
$\Delta t=0.1 fm$ is enough to get the convergency of differential equation Eq.\ref{Eq:discr}.
Within the numerical approach used in this paper both particle number and energy are conserved to an accuracy better than $10^{-4}$ within the time range explored in the following figures. We have also checked that
at the thermalization time $\tau_{eq.}$, the distribution reaches the equilibrium condition defined by Juttner-Boltzmann solution and the integral over distribution function is conserved at each time step.

In soft scattering approximation, another standard approach used to describe the HQ propagation in the bulk medium of quarks and gluons is by means of a Fokker-Planck equation of Eq.\ref{fokker-planck}. 
The Fokker-Planck equation is solved by a stochastic differential equation given by the Langevin equation where the equations of motion of the HQs are given by
\begin{eqnarray}\label{Langevin}
dx_i&=&\frac{p_i}{E}dt \nonumber \\
dp_i&=&-Ap_idt+C_{i,j}\rho_j\sqrt{dt}.
\end{eqnarray}
This set of equations describe the variation of coordinate $dx_i$ and momentum $dp_i$ in each time step $dt$ \cite{Rapp:2009my,LV,LV1}.
In the previous equation, $A$ represents drag force and $C_{i,j}$ is the covariance matrix that describes stochastic force in term of independent Gaussian-normal distributed random variables $\rho_j$. The random variable $\rho_j$ obey to the following distribution $p(\rho)=(2\pi)^{-3/2}e^{-\rho^2/2}$ with the conditions that $<\rho_i>=0$ and $<\rho_i \rho_j>=\delta(t_i-t_j)$.
This covariance matrix is related to diffusion coefficient in the following way:
\begin{equation}\label{C}
C_{i,j}=\sqrt{2B_T}P_{i,j}^\perp+\sqrt{2B_L}P_{i,j}^{||}
\end{equation}
where $B_T$ and $B_L$ are respectively the transverse and longitudinal component of diffusion coefficient. In general  $B_L=B_T=D$ for $p\rightarrow 0$ and it is a standard choice by several groups also at finite momenta $p$ when studying the HQ observables in realistic simulation of ultra-relativistic collisions \cite{LV,LV1,Rapp1,Hees}. In Langevin approach, the fluctuation-dissipation relation $B_T=TEA$ is commonly employed even if a microscopic derivation in general violates such relation at finite momentum as we have discussed in the previous sections. We have verified that the numerical solution of Langevin equation converges to the equilibrium solution $f_{eq}=e^{-E/T}$ at very large time. In order to fulfill this condition we reformulate the fluctuation-dissipation theorem as suggested by the pre-Ito interpretation\cite{Rapp:2009my} and we solve the Langevin equation with the condition:
\begin{equation}\label{pre-ito}
A(p)=\frac{D(p)}{ET}-\frac{D^\prime(p)}{p}
\end{equation}
therefore taking  $D(p)=B_T(p)$ as calculated by scattering matrix according to Eq. \ref{B1_comp} and Eq. \ref{F_finale} and we evaluate the correct drag force to achieve equilibrium distribution at themalization time. This procedure is necessary to guarantee that for $t\rightarrow\infty$ ($\gtrsim \tau_{eq.}$) 
also the Langevin approach converges to the correct equilibrium distribution as naturally occurs for the Boltzmann evolution.
Such agreement is shown in the right-low panel of Fig. \ref{Fig:OnOff}.

\subsection{Results on HQ moment evolution}
\begin{figure}[t]
\begin{center}
	\includegraphics[width=\linewidth]{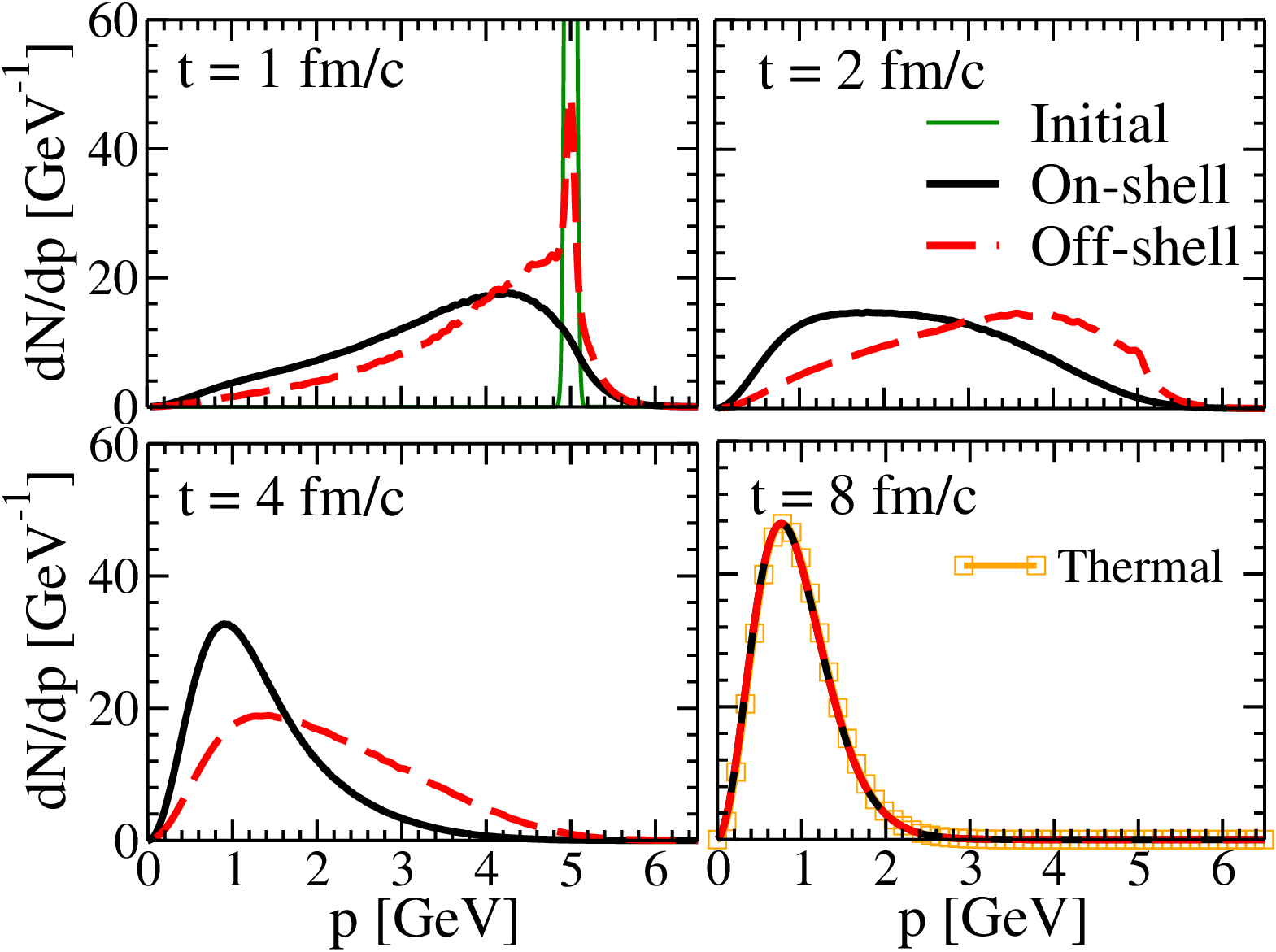}
	\caption{Charm quark momentum distribution as a function of the charm quark momentum at four different times $t=1 \, fm/c$ 
(left upper panel) and $t=2 \, fm/c$ (right upper panel), $t=4 \, fm/c$ (left lower panel) and $t=8 \, fm/c$ (right lower panel). 
Black solid lines are for on-shell dynamics while red dashed lines are for off-shell dynamics. 
The solid thin green line is the initial distribution that is the same for both calculations.}\label{Fig:OnOff}
\end{center}
\end{figure}
To investigate the differences between the heavy quark dynamics implied by Boltzmann 
on-shell dynamics and Off-shell dynamics, we study the heavy quark time evolution of the 
momentum distribution.
In the following results we have considered a thermal bulk of light quarks and
gluons at a temperature of $T=0.2 \, GeV$. In our calculation the initial charm quark distribution is assumed as an approximately 
delta distribution at $p_0 = 5 \,GeV$ shown by the green line in the left panel of Fig.\ref{Fig:OnOff}. 

In Fig. \ref{Fig:OnOff}, we show the time evolution of the momentum distribution $dN/dp$ for
both on-shell (black solid line) and off-shell dynamics (red dashed line) with $\gamma^*_{DQPM}$.
As shown the Boltzmann approach with off-shell collision integral has a slower dynamics than the on-shell one.
This can be understood as due to the fact that in the off-shell case the spreading of the bulk mass according to the quarks
and gluons spectral functions can be assimilated as a system with a larger average effective mass,
considering that the part of the spectral function at larger mass has anyway a larger phase space. 
At $t > 4 \, fm/c $ for both cases the momentum distribution tend towards a thermal distribution at $T=0.2 \, \rm GeV$ 
as shown in the right lower panel of Fig.\ref{Fig:OnOff} by the open square points.
The main difference is a faster evolution for the on-shell case that is however
mainly due to the fact that the on-shell and off shell dynamics have an underlying bulk system with a different
energy density and the drag coefficients are those corresponding to Fig.\ref{Fig:drag_eps}

In the following discussion we will show instead two different calculations for
two different drag and diffusion coefficient implementation.
The motivation is twofold. From one hand we try to discard the pure
off-shell effect in the HQ dynamics from the on-shell one. From the
other hand we are motivated by the fact that different approaches have
been used to extract the HQ transport coefficients from the comparison
of the observables, like nuclear modification factor and anisotropic flows,
with the experimental data.
In particular we will compare the results obtained within Langevin approach and within 
the on-shell and off-shell dynamics.
We firstly have considered one case where we scale the drag
coefficient of the off-shell kernel to the on-shell one by the energy density for the case
at larger width considered $\gamma^*/M=0.75$.

\begin{figure}[t]
	\begin{center}
		\includegraphics[width=\linewidth]{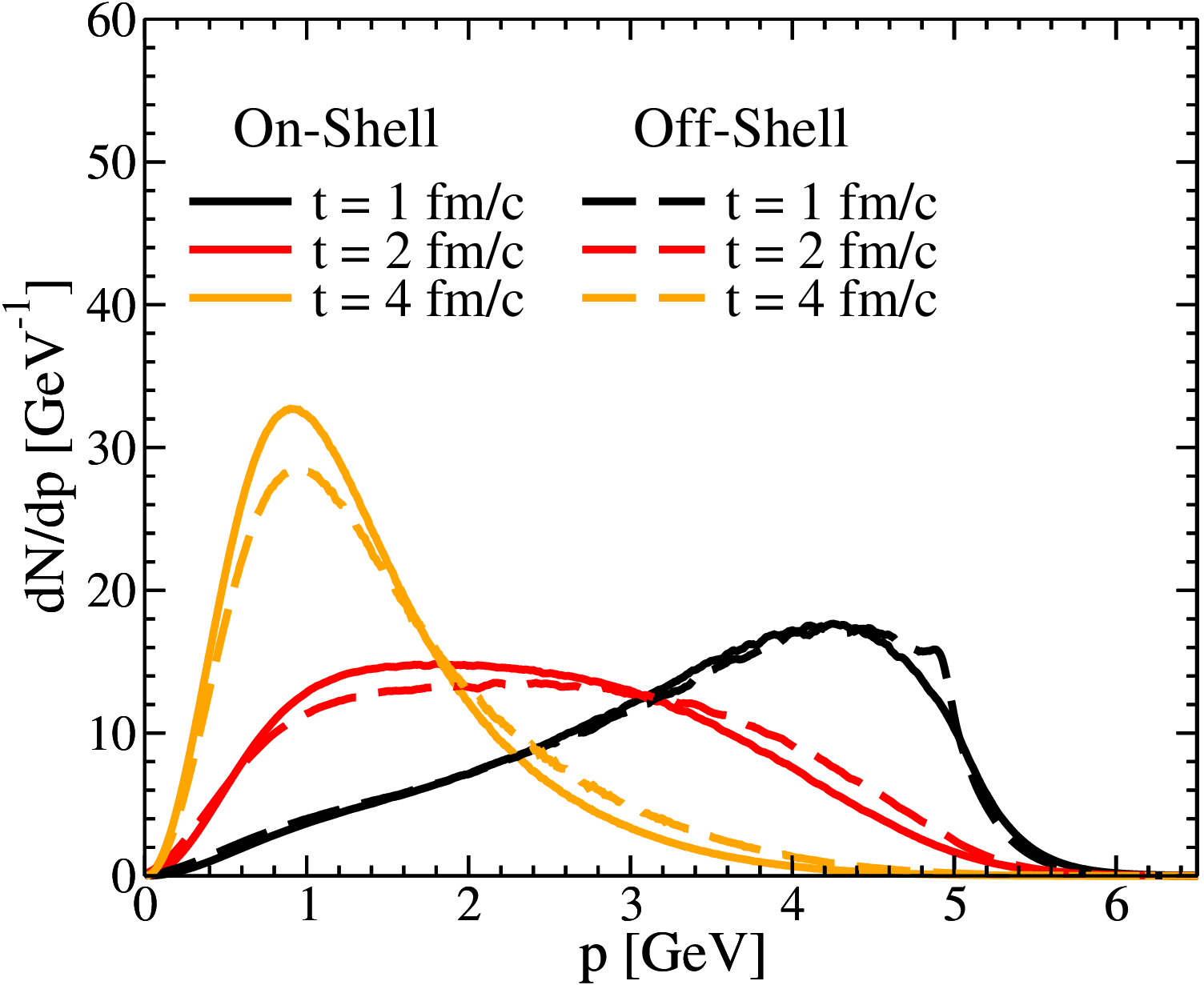}
	        \caption{Time evolution of the charm quark momentum distribution in a thermal bulk at $T=0.2 \, GeV$. Solid lines are for the on-shell dynamics while dashed lines are for off-shell dynamics. Different colors are for different times. For the off-shell case in this calculation $\gamma^*_i/M_i=0.75$ and the energy density of the bulk is the same to the on-shell case.}\label{Fig:evo_den}
\end{center}
\end{figure}
Here we considered the evolution on-shell and off-shell ($\gamma^*/M=0.75$) for a bulk
that has been tuned to have the same energy density in agreement with lQCD calculation.
This case is different from the previous one because here the bulk QGP as the same energy density
in the two cases.
As was shown in the previous section in Fig.\ref{Fig:drag_eps}, the effect of
the transport coefficient between off-shell and on-shell in large part due to
the difference in the energy density is damped when considered the physical case
where both on-shell and off-shell are tuned to the same energy density. 
In order to achieve this point, we upscale the off-shell scattering matrix by a constant factor
$k=\epsilon_{on-Shell}(T)/\epsilon_{off-Shell}(T)$ that in our simulation for a thermal bulk
at $T=0.2 \, GeV$ is about $k\approx 1.5$ corresponding to an underlying increase of the
coupling $g(T)$ of about a $6\%$.
In Fig.\ref{Fig:evo_den} it is shown the time evolution of the charm momentum distribution for both 
on-shell and off-shell dynamics.
We can notice that off-shell drag coefficient remains smaller than on-shell one especially at low momenta
and this implies an off-shell dynamics that is slightly slower with respect to the on-shell case,
but the effect remains quite small.
From these calculation we can assert that the differences seen in Fig. \ref{Fig:OnOff} are mainly due to
the different energy density induced by the fact that keeping equal the pole value of the mass and dressing
the system by a finite width induces a decreasing of the energy of the system.
Such an effect is nearly negligible for the off-shell case with $\gamma^*_{DQPM} \simeq 0.3-0.4\, M$, but becomes sizeable for
$\gamma^*/M=0.75$. However the
 pure off-shell dynamics does not show relevant differences as we can see comparing the on-shell (solid lines) 
 and off-shell mode (dashed lines). 
This suggests that on-shell Boltzmann equation is still a quite good approximation to study the evolution
of the charm momentum distribution at least up to $\gamma^*<M$.
\begin{figure}[t]
	\begin{center}
		\includegraphics[width=\linewidth]{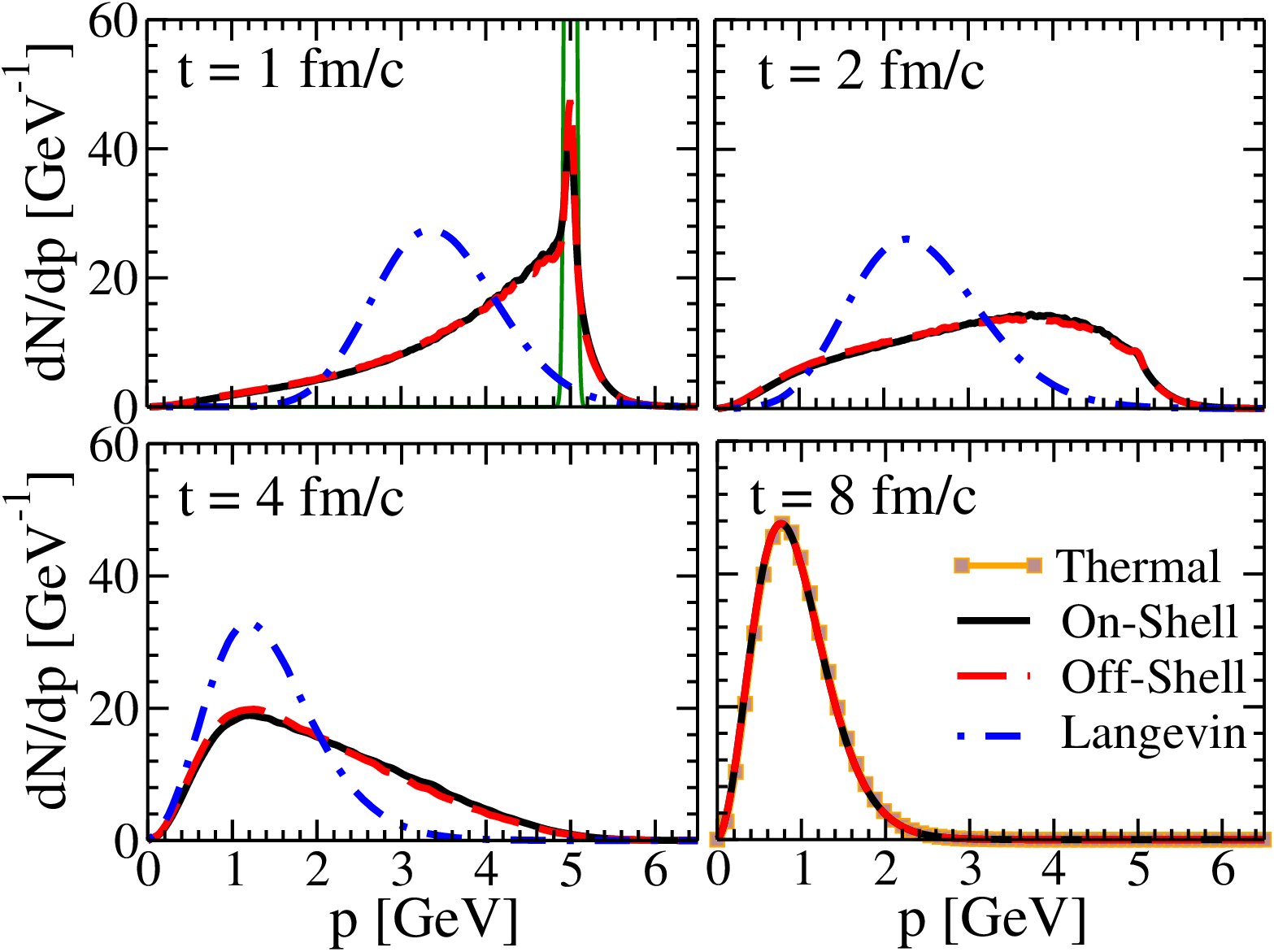}
		\caption{Charm quark momentum distribution as a function of the charm quark momentum at four different times $t=1 \, fm/c$ 
			(left upper panel) and $t=2 \, fm/c$ (right upper panel), $t=4 \, fm/c$ (left lower panel) and $t=8 \, fm/c$ (right lower panel). 
			Black solid lines are for on-shell dynamics, red dashed lines are for off-shell dynamics while blue dash-dotted lines are for langevin dynamics. 
			The solid thin green line is the initial distribution that is the same for each calculations.}\label{Fig:OnOff_DRAG2}
	\end{center}
\end{figure}

Finally, we have performed another calculation where we upscale the on-shell scattering matrix $|{\cal M_{Q}}|^2$
in order to reproduce the same drag coefficient obtained with the off-shell collision integral. This corresponds to 
multiply the on-shell scattering matrix $|{\cal M_{Q}}|^2$ by a function $k(p)$.
It may be considered as non realistic case because we have seen that the impact of off-shell dynamics on transport
coefficient is momentum dependent and leads to induce a slower increase of the drag coefficient at lower momenta.
We have considered it to study theoretically what happens if the interaction is such to generate exactly
the same drag $A(p)$ at each momentum. We show the results of this set-up for the case $\gamma_{DQPM}$.
in Fig. \ref{Fig:OnOff_DRAG2}. We can see that the time evolution of the HQ momentum distribution for the three different approaches on-shell Boltzmann (solid lines), off-shell Boltzmann (dashed lines) and Langevin (dot-dashed lines). 
By comparing solid lines and dashed lines once where one impose the same drag coefficient the two approaches show the same evolution.
Notice that in the Langevin calculations, we have used the pre-ito prescription
where the diffusion coefficient is the one obtained within off-shell
calculation shown in Fig.\ref{BT200}.
As shown the Langevin dynamics consists of a shift of the average momenta
with a fluctuation around it. This include the possibility that HQ
obviously lose energy moving the distribution to lower momenta but at
the same time they can gain energy from the bulk producing a tail with
momentum larger than the initial HQ momentum $p_0$. 
As shown, by the comparison between the Langevin and Boltzmann dynamics, 
the Boltzmann evolution of the charm quarks momentum 
does not have a Gaussian shape like one for the Langevin approach.
Where at the initial time the Boltzmann dynamics with respect 
to the Langevin one shows at initial stages a larger 
contribution from the gain term in the collision integral
with a global shape that is far from the Gaussian shape 
typical of Brownian motion \cite{Scardina:2014gxa}.

\subsection{Nuclear Modification factor $R_{AA}$ in Boltzmann and off-shell dynamics}
One of the main HQ observable investigated at RHIC and LHC energies is Nuclear Modification factor $R_{AA}$. It expresses the effective energy loss in Nucleus-Nucleus collision with respect to the production in proton-proton collisions.
In general, $R_{AA}$ gives a quantitative estimate of heavy quarks-bulk interaction.
Motivated by the phenomenological point of view, we have studied the impact of the results shown in the previous section on the evolution of the spectra in terms of the
$R_{AA}(p)$ for charm quarks.
We evaluate the Nuclear Modification factor using the charm quark distribution function at $t=0$ and $t=t_f$ as $R_{AA}=f_C(p,t_f)/f_C(p,t_0)$ both for on-shell and off-shell dynamics.
In these calculation for the initial momentum distribution of charm quark, we have used the charm quark production in Fixed Order + Next - to - Leading Log (FONLL)~\cite{Cacciari:2012ny} which describes the D-meson spectra in proton-proton collisions after fragmentation. 
\begin{figure}[t]
	\begin{center}
		\includegraphics[width=\linewidth]{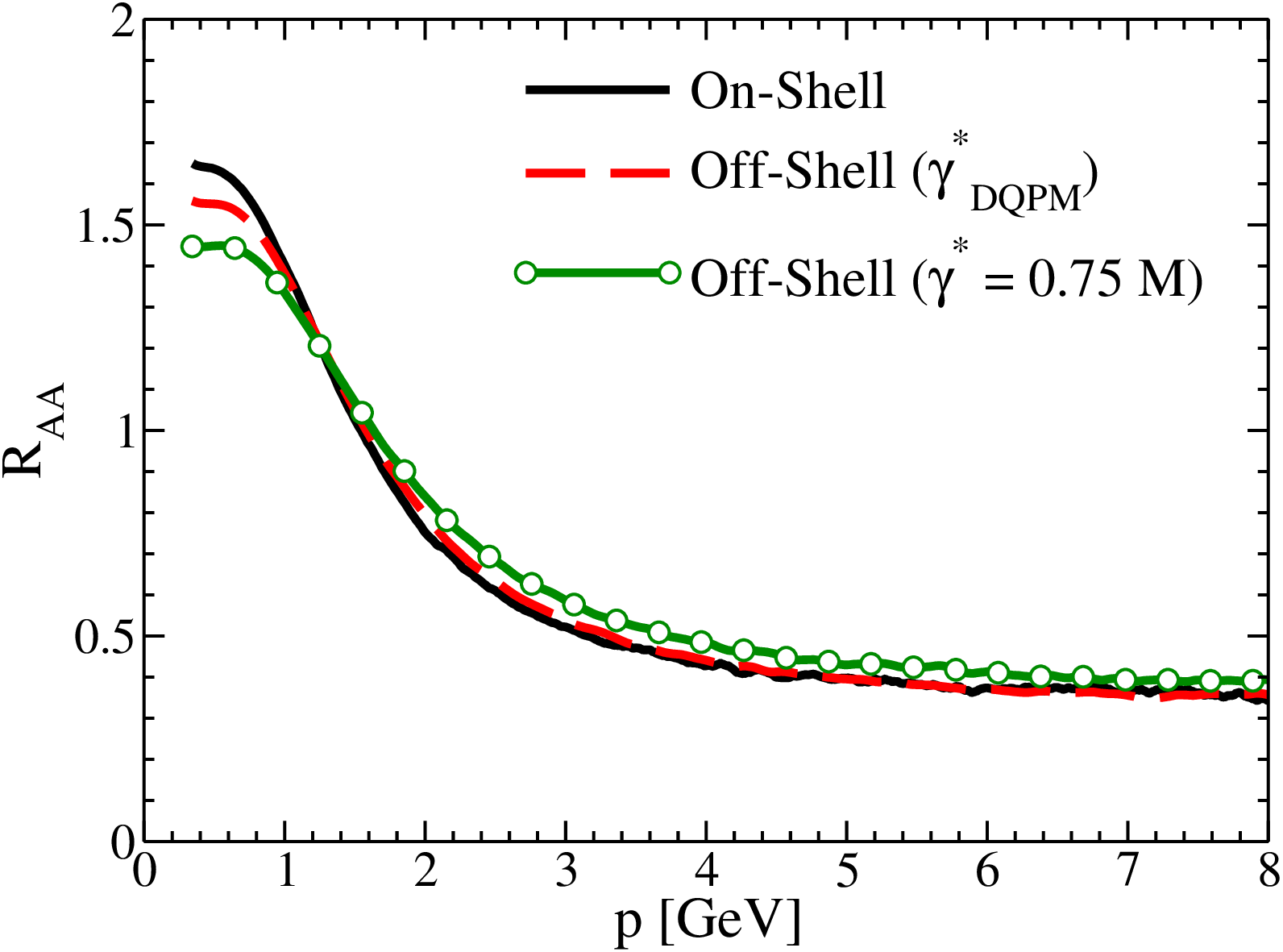}
	        \caption{Nuclear modification factor, $R_{AA}$ as function of charm momentum $p$. Solid lines refers to the case of on-shell calculations. The red dashed line refer to off-shell calculation with scattering matrix scaled with the energy density while the green open circles refer to the same calculation but with larger width fixed to $\gamma^*_i/M_i=0.75$.}\label{Ra}
\end{center}
\end{figure}

In Fig.\ref{Ra} we show the nuclear modification factor $R_{AA}$ as a function of the charm quark momentum for both 
on-shell and off-shell. Black Solid line refers to the case of on-shell calculations while the red dashed line and the green open circles refer to off-shell calculation with scattering matrix scaled to the case of the on-shell energy density and with $\gamma^*_{DQPM}$ and $\gamma^*_i/M_i=0.75$ respectively.
As shown, by comparing solid and dashed line, in the off-shell approach with $\gamma^*_{DQPM}$ the nuclear modification 
factor $R_{AA}$ does not show significant difference with respect to on-shell calculations especially for 
intermediate and high momentum of quark charm 
as shown in the same condition for evolution of distribution function. Also for the case of off-shell dynamics 
with $\gamma^*/M_i=0.75$ we find that the $R_{AA}(p)$  is slightly larger at high $p$ than on-shell one and it differs from the on-shell calculation less than $10\%$. 
Therefore a main result of thi work is that the off-shell dynamics does not modify significantly
the relation between $R_{A,A}(p)$ and $D_s(T)$ and it would not 
represent a main source of uncertainty in the phenomenological determination
of the space diffusion coefficient that are currently more dependent on the hadronization
mechanism, Langevin versus Boltzmann transport equation, assumption for bulk QGP expansion,
effects of non-equilibrium in the initial stage \cite{Xu:2018gux,Cao:2018ews,Rapp:2018qla}.

\section{Summary and conclusion}\label{Section:5}

We have studied the impact of off-shell dynamics on the drag and diffusion transport coefficients. 
We have found that if one just include the off-shell dynamics of quasi-particles associated to a finite mass width
this induce a moderate decrease of the density of the system an this leads to a smaller
drag and diffusion charm coefficient that is dependent on charm momentum. 
However when the comparison is done renormalizing the energy density of the
system, that is the one of lattice QCD, one can see that the main effect of off-shell dynamics is to reduce the increase
of the drag $A(p,T)$ and $B_T(p,T)$  at lower momenta $p \lesssim 2-3 \, \rm GeV$. Such a reduction
depends of the width and is maximal at $p=0$ being for $\gamma_{DQPM} \approx 0.3-0.4 m_{q,g}$ 
about a $25\% $ while increasing up to about a $35\%$ for the  $\gamma^*< 0.75\, m_{q,g}$.
In both case at $p> 3 GeV$ such a difference disappears completely.

We then have studied how a charm of momentum $p$ loose energy in a bulk QGP in equilibrium at temperature T=0.2 GeV,
comparing for the first time the time evolution of the momenta in a Langevin, Boltzmann on-shell and Boltzmann off-shell
transport approach. We find that at least in the regime of widths $\gamma^*< M$ the evolution of charm momenta
are only slightly modified by off-shell dynamics, also the impact of the last on $R_{AA}(p)$ is of about a $5\%$
at least at momenta $p> 1\, \rm GeV$.
Therefore from a phenomenological point of view the relation between the nuclear modification factor $R_{AA}(p_T)$
and the space diffusion coefficient  (or the drag)  is not significantly modified by off-shell dynamics.

\vspace{2mm}
\section*{Acknowledgments}
S.P. , M.L.S., and V.G. acknowledge the support of INFN-SIM national project and linea di intervento 2, DFA-Unict.
S.P. , M.L.S., and V.G. acknowledge the stimulating discussions and comments with G. Coci.

\vspace{3mm}


\end{document}